\RequirePackage{fix-cm}
\documentclass[final]{svjour3}                     
\pdfoutput=1
\smartqed  
\usepackage{graphicx}

\usepackage{booktabs} 
\usepackage{multirow}
\usepackage[algoruled,norelsize]{algorithm2e} 
\usepackage{tablefootnote}

\usepackage{mathtools}

\usepackage{natbib}

\newcommand{\ignore}[1]{}

\DeclarePairedDelimiter{\ceil}{\lceil}{\rceil}

\begin{document}

\title{Evaluating Sentence-Level Relevance Feedback for High-Recall Information
Retrieval
}


\author{Haotian Zhang         \and
        Gordon V. Cormack        \and
        Maura R. Grossman   \and
        Mark D. Smucker
}


\institute{Haotian Zhang, Gordon V. Cormack, Maura R. Grossman \at
           David R. Cheriton School of Computer Science, University of Waterloo, Waterloo, ON, Canada \\
           \email{haotian.zhang@uwaterloo.ca}           
           \and
           Mark D. Smucker \at
           Department of Management Sciences, University of Waterloo, Waterloo, ON, Canada 
}


\maketitle

\begin{abstract}
This study uses a novel simulation framework to evaluate whether the
time and effort necessary to achieve high recall using active learning is reduced by presenting
the reviewer with 
isolated sentences, as opposed to full documents, for
relevance feedback.  Under the weak assumption that more time
and effort is required to review an entire document than a single sentence,
simulation results indicate that the use of isolated sentences
for relevance feedback can yield comparable accuracy and higher efficiency,
relative to the state-of-the-art Baseline Model Implementation (BMI) of
the AutoTAR Continuous Active Learning (``CAL'') method employed in the
TREC 2015 and 2016 Total Recall Track.

\end{abstract}

\section{Introduction}

There are several application domains---including legal e-discovery and
systematic review for evidence-based medicine---where finding all, or
substantially all, relevant documents is crucial.  Current
state-of-the-art methods for achieving high recall rely on
machine-learning methods that learn to discriminate between relevant and
non-relevant documents based on large numbers of human relevance
assessments. In many instances, thousands of assessments may be required.
These human assessments represent the primary cost of such methods,
which can be prohibitive when expert assessments are required.  In
this work, we examine whether it is possible to use sentence-level
assessments in place of document-level assessments to reduce the time
needed to make judgments, the number of judgments needed, or both.  We
present a novel strategy to evaluate this hypothesis, and
show simulation results using standard test collections indicating
that assessment effort can be reduced to judging a single
sentence from a document without meaningful reduction in
recall. Replacing documents with sentences has the potential to reduce
the cost and burden associated with achieving high recall in many
important applications.

Simulation methods have long been a staple of information-retrieval (IR) evaluation.
The dominant methodology of studies reported in the literature
derives from Sparck Jones and Van Rijsbergen's ``ideal'' test 
collection \citep{spark1975report}, in which the results of
ad hoc searches for each of a set of topics within a dataset are 
compared to relevance labels for a subset of the documents, rendered after the fact
by human assessors. This approach is generally considered to yield
reliable comparisons of the relative effectiveness ad hoc IR 
systems that do not rely on relevance feedback.

To simulate relevance feedback, we require a substantially complete set of relevance labels
prior to the simulation; the reviewer's response to any particular
document during the simulation is determined by consulting these previously determined
labels.  Furthermore, to simulate the presentation of 
isolated sentences rather than documents to the reviewer for feedback, we require a 
prior relevance label for each sentence in every document,
with respect to every topic.  

In the current study, we augment four publicly available test collections with 
sentence-level relevance labels derived using a combination of the available 
relevance labels, new assessments, heuristics, and machine-learning
(Section~\ref{section.SCAL}).  We use the
available labels to simulate document-level relevance feedback, and the newly
created labels to simulate sentence-level relevance feedback (Section~\ref{section.cal}).  Both are evaluated
in terms of document-level recall---the fraction of relevant documents presented
in whole or in part to the reviewer---as a function of reviewer effort.  Effort is measured
in two ways---as the total number of assessments rendered by the reviewer, and as the total
number of sentences viewed by the reviewer in order to render those assessments (Section~\ref{section.eval}).
We assume that the reviewer's actual time and effort is likely to fall somewhere between these two
bounds.

In addition to choosing whether to present a full document or isolated 
sentence to the reviewer for feedback, it is necessary to choose the manner in
which the document or sentence is selected.  As a baseline, we used the Baseline
Model Implementation (``BMI'') implementation of the AutoTAR Continuous Active Learning
method (``CAL'') shown in Section~\ref{section.related}, which repeatedly 
uses supervised learning to select and present to the reviewer for labeling the next-most-likely relevant
document, which is then added to the training
set. We extended BMI to incorporate three binary choices:  (1) whether to \emph{present}
full documents or sentences to the reviewer for feedback; (2) whether to \emph{train} the learning
algorithm using full documents or isolated sentences; and (3) whether to \emph{select} the highest-scoring
document, and the highest-scoring sentence within that document, or to select the highest
scoring sentence, and the document containing that sentence.  We evaluated all eight 
combinations of each of these three binary choices in Section~\ref{section.cal}.

We conjectured that while sentence-level feedback might be less accurate than document-level
feedback, yielding degraded recall for a given number of assessments, that
sentence-level feedback could be rendered more quickly, potentially yielding higher
recall for a given amount of reviewer time and effort. We further conjectured that selecting
the highest-scoring sentence (as opposed to the highest-scoring document) and/or 
using sentences (as opposed to documents) for training might help to improve
the accuracy and hence efficiency of sentence-level feedback.

Contrary to our conjecture, we found that sentence-level feedback resulted in
no meaningful degradation in accuracy, and that the methods intended to mitigate the
anticipated degradation proved counterproductive (Section~\ref{section.results}).  
Our results suggest that relevance
feedback based on isolated sentences can yield higher recall with less time
and effort, under the assumption that sentences can be assessed, on average,
more quickly than full documents.

\section{Related Work}
\label{section.related}
While the problem of \textbf{H}igh-\textbf{R}ecall-\textbf{I}nformation-\textbf{R}etrieval (HRIR) 
has been of interest since
the advent of electronic records, it currently commands only a small fraction of
contemporary IR research.  The most pertinent body of recent HRIR research derives from
efforts to improve the effectiveness and efficiency of Technology-Assisted Review (``TAR'') for electronic discovery (eDiscovery)
in legal, regulatory, and access-to-information contexts, where the need is to find all or
substantially all documents that meet formally specified criteria within a finite corpus.
A similar problem has been addressed within the context of systematic review for evidence-based
medicine and software engineering, where the need is to find reports of substantially
all studies measuring a particular effect. Constructing an ideal test collection 
for IR evaluation entails a similar need:  to identify substantially all of the relevant documents
for each topic. Although the focus of TREC has diversified since its inception in 1992, and
methods to achieve high-recall have evolved, the original impetus for TREC was to support
the needs of information analysts, who ``were willing to look at many documents and 
repeatedly modify queries in order to get high recall.'' \citep{voorhees2005trec}.

The method of conducting multiple searches with the aim of achieving high recall,
dubbed \textbf{I}nteractive \textbf{S}earch and \textbf{J}udging (ISJ), while common, has rarely been evaluated
with respect to how well it achieves its overall purpose.  The initial TREC tasks
evaluated one single search, assuming that improvements would contribute
to an end-to-end process involving multiple searches.  
An early study by \citet{blair1985evaluation} indicated that searchers
employing ISJ on an eDiscovery task believed they had achieved 75\% recall
when in fact they had achieved only 20\%.  Within the context of the TREC 6 ad hoc task,
\citet{cormack1998efficient} used ISJ to achieve 80\% recall with 2.1 hours of effort, on average, for
each of 50 topics. A principal difference between the two studies is that Cormack et al. used
``shortest substring ranking and an interface that displayed relevant passages and 
allowed judgments to be recorded,'' whereas Blair and Maron used Boolean searches
and reviewed printed versions of entire documents.

The current states of the art for HRIR and for its evaluation are represented by
the tools and methods of the TREC Total Recall Track, which ran in 2015 and 2016 \citep{roegiest2015trec,roegiest2016trec},
and form the baseline for this study.  The Total Recall Track protocol simulates
a human in the loop conducting document-level relevance assessments, and measures recall as
a function of the number of assessments, where recall is the fraction of all relevant
documents presented to the reviewer for assessment. BMI, an HRIR implementation conforming
to the Total Recall protocol, was supplied to Total Recall Track participants in advance, and used as the
baseline for comparison.  

\begin{figure}
\centering
\includegraphics[width=0.75\textwidth]{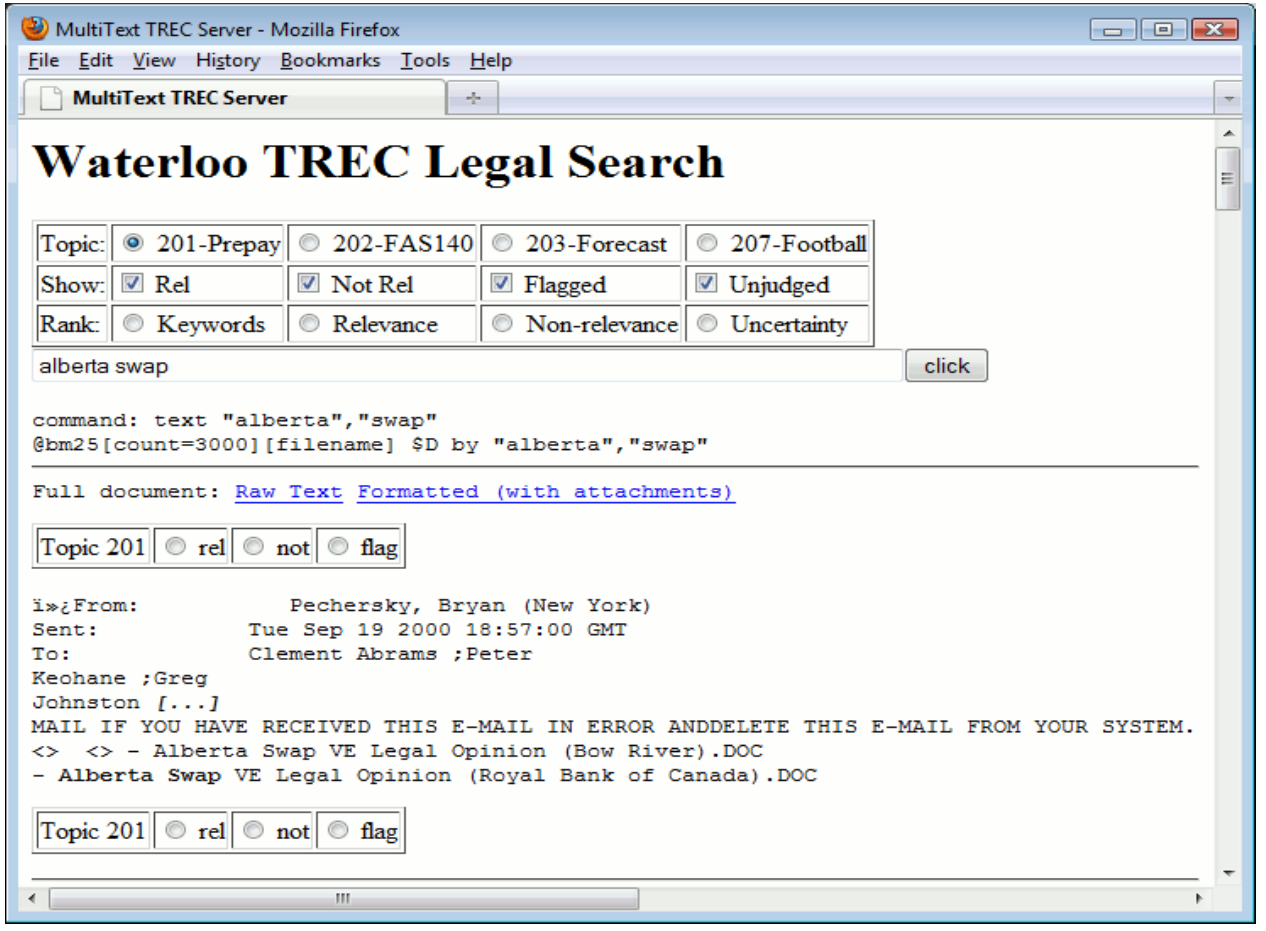}
\caption{The result page presented by Cormack and Mojdeh's ISJ tool \citep{cormack2009machine}.}
\label{irj-ui}
\end{figure}

No method evaluated in the TREC Total Recall Track surpassed the overall effectiveness of BMI 
\citep{roegiest2015trec,roegiest2016trec,DBLP:conf/trec/ZhangLWCS15}.  
A prior implementation of the same method had been shown to surpass
the effectiveness of the ISJ results of \citep{DBLP:journals/corr/CormackG15} on the TREC 6 
data shown in Figure~\ref{irj-ui}, as well as a similar method 
independently contrived and used successfully by \citet{soboroff2003building} to construct 
relevance labels for the TREC 11 Filtering Track \citep{robertson2002trec}. Recently, BMI and a
method independently derived from CAL have produced results that compare favorably to competing
methods for systematic review \citep{kanoulas2017clef,CormackG17Clef,baruah2016optimizing}.  BMI has shown effectiveness
that compares favorably with exhaustive manual review in categorizing 402,000 records
from Governor Tim Kaine's administration as Governor of Virginia \citep{cormack2017navigating}.

BMI is an implementation of CAL, which is effectively
a relevance-feedback (RF) method, albeit with a different objective and implementation
than to construct the ultimate query by selecting and weighting search terms, as 
typically reported in the RF literature 
\citep{aalbersberg1992incremental,ruthven2003survey}.
CAL uses supervised machine-learning 
algorithms that have been found to be effective for text categorization, but with the 
goal of retrieving every relevant document in a finite corpus, rather 
than to construct the ultimate automatic classifier for a hypothetical infinite 
population.  Given these differences, results from RF and text categorization 
should not be assumed to apply to CAL. In particular, relevance feedback for non-relevant
documents has been shown to be important for CAL \citep{DBLP:conf/trec/PickensGHN15}, 
while uncertainty sampling
has shown no effectiveness benefit over relevance sampling, while incurring added 
complexity \citep{cormack2014evaluation}.

The TREC Legal Track (2006--2011) 
\citep{baron2006trec,tomlinson2007overview,oard2008overview,hedin2009overview,cormack2010overview,grossman2011overview} 
investigated HRIR methods for eDiscovery, which
have come to be known as TAR.  The main task from 2006 through 2008
evaluated the suitability of ad hoc IR methods for this task, with unexceptional
results. A number of RF and text categorization tasks were also posted,
each of which involved categorizing or ranking the corpus based on a fixed set of
previously labeled training examples, begging the question of how this training set
would be identified and labeled within the course of an end-to-end review effort starting
with zero knowledge. 2008 saw the introduction of the interactive task, 
reprised in 2009 and 2010, for which teams
conducted end-to-end reviews using technology and processes of their own choosing,
and submitted results that were evaluated using relevance assessments on a non-uniform
statistical sample of documents.  In 2008 and 2009, San Francisco e-discovery service provider H5 achieved superior results using
a rule-based approach \citep{hogan2008h5}; 
in 2009 the University of Waterloo employed 
a combination of ISJ and CAL to achieve comparable results \citep{cormack2009machine}.
In a retrospective study using secondary data from TREC 2009 \citep{grossman2010technology}, two of the authors of the current study
concluded that the rule-based and ISJ+CAL approaches both yielded results that compared favorably
to the human assessments used for evaluation.  It was not possible, however, given the design
of the TREC task, to determine the relative contributions of the technology, the process,
and the quality and quantity of human input to the H5 or Waterloo results.

Prior to CLEF 2017, the systematic review literature described primarily text categorization
efforts similar to those employed by the TREC Legal Track, in which the available data were 
partitioned into training and test sets, and effectiveness evaluated with respect to
classification or ranking of the test set \citep{hersh2003trec,wallace2010active,wallace2013modernizing}.  One notable 
exception is \citet{yu2016read} which affirms the effectiveness of CAL 
for systematic review.

Contemporary interactive search tools---including the tools employed for ISJ---typically
display search results as document surrogates \citep{hearst2009search}, which consist of excerpts or summaries from which the reviewer can decide whether
or not to view a document, or whether or not to mark it relevant.  For example, the ISJ method
described above used the result rendering shown in Figure~\ref{irj-ui}, which consists of a fragment
of text from the document, accompanied by radio buttons for the reviewer to render a relevance assessment.
Typically, the surrogate consists in whole or in part of a query-biased summary or excerpt of
the full document.

\citet{tombros1998advantages} found that reviewers could identify more relevant documents for each query 
by reviewing the extracted summary, while, at the same time, making fewer labeling errors. 
In a subsequent study, \citet{sanderson1998accurate} found that ``[t]he results reveal that reviewers can 
judge the relevance of documents from their summary almost as accurately as if 
they had had access to the document's full text.''
An assessor took, on average, $24$ seconds to assess each summary and $61$ seconds to assess each full 
document.
\citet{smucker2010human} also used query-biased snippets of documents for relevance 
judgment in a user-study setting. The results show that the average time to judge a summary was around $15.5$ 
seconds while the time to judge a document was around $49$ seconds. Smucker and Jethani also found that 
reviewers were less likely to judge summaries relevant than documents.

In passage retrieval, the goal is to accurately identify fragments of 
documents,---as opposed to entire documents---that contain relevant information. 
Some studies \citep{allan2005hard,salton1993approaches} have shown that passage retrieval
can help to identify relevant documents and hence to improve
the effectiveness of document retrieval.
\citet{liu2002passage} used passage retrieval in a language model and found that passages can 
provide more reliable retrieval than full documents.

To evaluate the effectiveness of passage retrieval systems, the TREC 2004 HARD Track
employed an adapted form of test collection, in which assessors were asked to partition
each relevant document to separate regions of text containing relevant information from
regions containing no relevant information.  

The accuracy and completeness of relevance assessments in test collections has been
an ongoing concern since their first use in IR evaluation \citep{voorhees2005trec}.
It is well understood that it is typically impractical to have a human assessor label every 
document in a realistically sized corpus; it is further understood that human assessments
are not perfectly reliable.  Nonetheless, it has been observed that it is possible to 
select enough documents for assessment, and that human assessment is reliable enough to
measure the relative effectiveness of IR systems, under the assumption
that unassessed documents are not relevant.  The pooling method
suggested by \citet{spark1975report} and pioneered at 
TREC \citep{voorhees2005trec} appears to yield assessments that are
sufficiently complete---given the level of assessment effort and the size of the initial
TREC collections---to reliably rank the relative effectiveness of different IR systems.  
Other methods of selecting documents for
review, including ISJ, have been observed to be similarly effective, while entailing less
assessment effort \citep{cormack2009machine,soboroff2003building}.

Evaluation measures have been proposed that avoid the assumption that unassessed
documents are not relevant, gauging system effectiveness only on the basis of documents
for which relevance labels are available such as bpref \citep{clarke2005trec}.
\citet{Buttcher:2007:RIR:1277741.1277755} achieved reliable evaluation results by using an SVM classifier 
to label all of the unlabeled documents in TREC GOV2 collection \citep{buttcher2006trec}, using the labeled documents as a training set.
Systems were then evaluated assuming both the human-assessed and machine-classified labels
to be authoritative.

The problem of evaluating user-in-the-loop systems has been investigated using human
subjects as well as simulated human responses \citep{voorhees2005trec}.  For a decade, experiments using human subjects were
the subject of the TREC Interactive Track and related efforts \citep{over2001trec}, which
exposed many logistical hurdles in conducting powerful, controlled, and realistic
experiments to compare system effectiveness \citep{interactive-trec-track-putting-user-search, voorhees2005trec}.  Not
the least of these hurdles was the fact that the human subjects frequently disagreed
with each other, and with the labels used for assessment, raising the issue of how
the results of different subjects should be compared, and how human-in-the-loop
results should be compared to fully automated results.
To the authors' knowledge, no
controlled end-to-end evaluation of the effectiveness of HRIR 
methods has been conducted using human subjects.  

Simulated responses are more easily controlled, at the expense of realism.
The simplest assumption is that the human is infallible, and will assess a document
exactly as specified by the relevance label in the test collection.  This assumption
was made in relevance-feedback studies by \citet{drucker2001relevance}, the TREC Spam Track \citep{cormack2005trec},
and the TREC Total Recall Track \citep{roegiest2015trec,roegiest2016trec}. \citet{cormack2014evaluation} used a 
``training standard'' to simulate relevance feedback, separate from the ``gold standard''
used to estimate recall. \citet{cormack2017navigating} used the results of ``secondary''
assessors from TREC 4 to simulate feedback, separate from the primary assessor whose
labels are used to evaluate recall.  In the same study, Cormack and Grossman used assessments
rendered previously by the Virginia Senior State Archivist to simulate relevance feedback,
and post-hoc blind assessments by the same archivist to estimate recall.  Cormack and Grossman
distinguish between ``system recall,'' which denotes the fraction of all relevant documents
presented to the reviewer, from ``user recall,'' which denotes the fraction of all relevant documents
that are presented to the reviewer and assessed as relevant by the reviewer.

A second simplifying assumption in simulation experiments is to quantify reviewer effort
by the number of documents or surrogates presented to the reviewer for review.  
However, in eDiscovery industry, the human assessors are usually paid based on the total assessment time.
Therefore, the review speed of assessor can influence the review effort.
The reviewer's speed depends on a number of factors. \citet{RahbariaslShahin2018}
studied the effects of time constraints and document excerpts on the speed of relevance judgments.
In Rahbariasl's study, users were shown either full documents
or document excerpts. They were required to judge these documents within 15, 30, or 60 seconds time constraints.
Rahbariasl found that time constraints can increase the judging speed rate of assessors while did not hurt
the quality of judgments.
\citet{maddalena2016crowdsourcing} also reported that applying time constraints on assessments
would not lead to the loss of judgment quality.
\citet{wang2010user} evaluated the effects of different parameters towards relevance 
assessment. They found there was no significant difference in the assessment speed between
different groups of assessors. But the assessment speed varied for individuals.
In a followed up study conducted by \citet{wang2011accuracy}, Wang studied 
a number of influencing factors, such as document subjects, length, and legibility,
assessors reading skills and subject knowledge, relevance guidelines, and learning effects.
The results indicated that strong correlation was observed between perceived difficulty 
and assessment speed. Some difficult documents took noticeably longer time for assessors to review.
The document length was also a factor that influenced assessor's speed.
The review speed also varied significantly between different topics.


The challenge of acquiring a complete set of labels for relevance assessment was addressed 
within the context of the TREC Spam Track \citep{cormack2005trec}.
The Track coordinators used an iterative process \citep{cormack2005spam} in which
a number of spam classifiers were applied to the corpus, and disagreements between the classifiers
and a provisional labeling were adjudicated by the coordinators. The process was repeated several
times, until substantially all labels were adjudicated in favor of the provisional gold standard.
At this point, the provisional gold standard was adopted as ground truth, and its
labels were used to simulate human feedback and, subsequently, to measure effectiveness.  
A later study by \citet{kolcz2009genre} measured
the error rate of the gold standard, according to the majority vote of a crowdsourced labeling effort. 
The observed error rate for the gold standard---1.5\%---was considerably lower than the 
the observed error rate of 10\% for individual crowdsource workers.

The TREC 11 Filtering Track
coordinators used a method similar to CAL to identify documents which were assessed for relevance;
after TREC, further documents selected using the pooling method were assessed for relevance, and used to 
estimate the recall of the original effort to have been 78\% \citep{sanderson2004forming,DBLP:journals/corr/CormackG15}.  
The additional assessments did not materially affect the evaluation results.  

Cormack and Grossman \citep{cormack2014evaluation} found CAL to be superior to classical supervised learning and
active learning protocols (dubbed ``simple passive learning'' (SPL) and
``simple active learning'' (SAL), respectively) for HRIR. Cormack and Grossman observed
comparable results for TREC Legal Track collection detailed above, as well as four private
datasets derived from real legal matters.

The TREC Total Recall Track used a total of seven test collections \citep{roegiest2015trec,roegiest2016trec}.  
For five of the collections, including the collections used in the current study, the
Track coordinators used ISJ and CAL with two different feature engineering techniques and two different
base classifiers to identify and label substantially all relevant documents prior to running the task. 
These labels were used to simulate reviewer feedback and to evaluate the results.  For the 2016 Track,
an alternate gold standard was formed by having three different assessors label each of a non-uniform
statistical sample of documents for each topic \citep{roegiest2016trec,zhang2016sampling}.  The alternate assessments yielded substantially
the same evaluation results as the full gold standard.  Subsequently, \citet{cormack2017navigating} used
a revised gold standard for both simulation and evaluation, and found no material difference in
results.

Relevance labels for the TREC 2004 HARD Track, which were not used at the time to simulate reviewer feedback, 
but which are used for that purpose in the current study, were rendered for a set of documents
selected using the classical pooling method.

In a recent study from \citet{zhang2018cikm}, they conducted a controlled 50-users study to evaluate
using document excerpts (a single extracted paragraph from document) 
as relevance feedback in continuous active learning.
Participants were asked to find as many relevant documents as possible within one hour using the
HiCAL system~\citep{DBLP:conf/sigir/AbualsaudGZSCG18,zhang2017trec,zhang2018trec}.
They found that study participants were able
to find significantly more relevant documents within one hour when they used the
system with showing document excerpts (paragraphs) as opposed to full documents.

\section{Method}
\label{section.method}
This study mainly addresses the question of whether and
how sentence-level relevance feedback can achieve
high recall. Furthermore, in a given amount of review effort,
how much faster sentence-level relevance feedback is able to achieve a certain high recall
than document-level relevance feedback.

To investigate this question, we apply an extended version of BMI to augmented versions
of four public test collections, so as to simulate eight variants of sentence-level and 
document-level feedback. The eight variants varies on three different binary choices.
The first choice is to present sentence or document to assessor for relevance feedback.
The second choice is to add the reviewed sentence or the reviewed document into the training
set to retrain the machine learned classifier. The third choice is to rank on
all sentences or all documents in order to select the most relevant sentence or document for assessor to review.
By varying the choices on these three dimensions, we can compare different relevance feedback
strategies and derive the most effective strategy to achieve high recall.

As for the comparison of different strategies, we compare the recall achieved by these eight different
strategies at a given amount of effort. As mentioned in Section~\ref{section.related},
the effort to review a document and a sentence can be different.
To compare different relevance feedback strategies comprehensively, we apply
two different methods to model the review effort.
First, we simply count the number of assessments 
rendered by the simulated reviewer to achieve a certain recall. In this case,
the effort to review a document and the effort to review a sentence will
be the same. 
Second, we also count the number of sentences viewed by the simulated reviewer in rendering those 
assessments. In this case, we assume that a long document containing multiple sentences 
will cost more review effort than a single sentence.
By applying these two evaluation methods, the review effort can be measured and simulated
from a perspectives. 

\subsection{Continuous Active Learning with Sentence-Level or Document-level relevance feedback}
\label{section.cal}
\begin{algorithm}[t]
\normalsize
Step 1. Treat the topic statement as a relevant document and add this document into the training set\;
Step 2. Set the initial batch size $B$ to 1\;
Step 3. Temporarily augment the training set by adding 100 random documents from the collection, temporarily labeled ``non-relevant''\;
Step 4. Train a logistic regression classifier using the training set\;
Step 5. Remove the random documents added in Step 3 from the training set\;
Step 6. Select the highest-scoring $B$ documents from the not reviewed documents\;
Step 7. Append the selected $B$ documents to system output. The system output records the list of documents
that have been selected by the classifier and labeled by the reviewer\;
Step 8. Review the selected $B$ documents, coding each as ``relevant'' or ``non-relevant''\;
Step 9. Add the labeled $B$ documents into the training set\;
Step 10. Increase $B$ by $\ceil{\frac{B}{10}}$\;
Step 11. Repeat steps 3 through 10 until a sufficient number of relevant documents have been reviewed.
\caption{The autonomous TAR (AutoTAR) algorithm}
\label{alg.autotar}
\end{algorithm}

BMI implements the AutoTAR CAL method \citep{DBLP:journals/corr/CormackG15}, 
shown in Algorithm~\ref{alg.autotar}. The topic statement is labeled as a
relevant document and $100$ randomly selected documents are labeled as ``non-relevant'' in the training set shown in Steps 1 and 3. 
A logistic regression classifier is trained on this training set in Step 4. The highest-scoring $B$
documents are selected from the not reviewed documents and appended to system output in Steps 6 and 7. The system output
records the list of the reviewed documents. The $B$ documents labeled by reviewer are then added to the training set in Step 9.
$100$ randomly selected documents coded as non-relevant in the training set are replaced by the newly selected $100$ random documents in Step 3 and 5.
The classifier is re-trained using the new training set. The classifier selects the next $B$ highest-scoring 
not reviewed documents for review in the new batch. This process repeats until enough relevant documents have been found.

We modified BMI to use either sentences or
documents at various stages of its processing.  As part of this
modification, we consider the document collections to be the union of
documents and sentences, and choose documents or sentences at each step,
depending on a configuration parameter.  For example, a single document of $100$
sentences becomes $101$ documents, where $1$ document is the original
document and the other $100$ documents are the document's sentences.

BMI uses logistic regression as implemented by
Sofia-ML\footnote{https://code.google.com/archive/p/sofia-ml/} as its
classifier.  The logistic regression classifier was configured with
logistic loss with Pegasos updates, L2 normalization on feature
vectors with $lambda = 0.0001$ as the regularization parameter, AUC
optimized training, and $200,000$ training iterations. The features
used for training the classifier were word-based tf-idf:
\begin{equation}
w = (1 + \log(\mathit{tf})) \cdot log(N/\mathit{df})
\end{equation}
where $w$ is the weight of the word, $\mathit{tf}$ is the term frequency, $N$
is the total number of documents and sentences, and $\mathit{df}$ is the
document frequency where both documents and sentences are counted as documents.
The word
feature space consisted of words occurring at least twice in the
collection and all the words were downcased and stemmed by the Porter
stemmer.  

\begin{algorithm}[t]
\normalsize
Step 1.  Treat the topic statement as a relevant document and add this document into the training set\;
Step 2.  Set the initial batch size $B$ to 1\;
Step 3.  Temporarily augment the training set by adding 100 random 
documents ($2d$) or sentences ($2s$) from the collection, temporarily labeled ``non-relevant''\;
Step 4.  Train the classifier using the training set. Then remove the random documents added in Step 3 from the training set\;
Step 5.  Derive the top $B$ (best\_sent, best\_doc) pairs using the classifier. We have two choices $\{3d,3s\}$ to
select the (best\_sent, best\_doc) pair. The details of the $\{3d,3s\}$ are shown in Table~\ref{combinations.table}\; 
Step 6.  Append the selected $B$ best\_doc to system output (coded as $O$). The system output records the list of best\_doc
that have been selected by the classifier and labeled by the reviewer\;
Step 7.  For each of the top $B$ (best\_sent, best\_doc) pairs execute steps 8 to 10\;
Step 8.  Present either the best\_sent ($1s$) or best\_doc ($1d$) in the pair to the reviewer\;
Step 9.  Receive the relevance assessment $l$ from reviewer\;
Step 10.  Add either (best\_sent, $l$) as $2s$ or (best\_doc, $l$) as $2d$ to training set\;
Step 11.  Increase $B$ by $\ceil{\frac{B}{10}}$\;
Step 12.  Repeat steps 3 through 11 until substantially all relevant documents appear in the system output.
\caption{Generic sentence feedback and document feedback algorithm}
\label{alg.gen}
\end{algorithm}

Algorithm~\ref{alg.gen} illustrates our modified BMI 
that enables either sentence-level or document-level feedback,
training, and ranking.  The system output in Step 6 records the documents
that have been labeled by reviewer. The system output also keeps the order of documents
judged by reviewer so that we can use the system output to measure the recall achieved at a certain amount of effort.

Steps 3, 5, 8 and 10 involve choices; we explored two possibilities for each
choice, for a total of eight combinations.  The principal choice
occurs in Step 8: whether to present to the reviewer the best\_sent or
the best\_doc in the pair.  We label these
alternatives $1s$ and $1d$, respectively.  In support of this choice,
it is necessary to choose how to build the training set in steps 3 and
10, and how to use the classifier to identify the top $B$
(best\_sent, best\_doc) pairs in Step 5. In Step 10, we choose as new
added training examples either: ($2s$) the best\_sent with
corresponding label $l$; or ($2d$) the best\_doc with corresponding
label $l$. In step 3, the 100 randomly selected non-relevant training
examples are chosen by either: ($2s$) 100 random sentences; or ($2d$)
100 random documents. In Step 5, we choose the (best\_sent, best\_doc)
pair either: ($3s$) the highest-scoring sentence contained in any
document not yet in system output, and the document containing that
sentence; or ($3d$) the highest-scoring document not yet in system
output, and the highest-scoring sentence within that document. The
sentences in ($3d$) were scored by the same classifier that was
also used for document scoring.  More formally, if we denote system
output by $O$, $3s$ is defined by Equations~\ref{best.doc.s} and
\ref{best.sent.s}:

\begin{equation} best\_sent =\underset{sent \notin doc \in O }{\mathrm{argmax}}  \  Score(sent) 
\label{best.doc.s}
 \end{equation}
   
\begin{equation} best\_doc = d \mid best\_sent \in d
\label{best.sent.s}
 \end{equation}

while $3d$ is defined by Equations~\ref{best.doc.d} and~\ref{best.sent.d}:

\begin{equation} best\_doc =\underset{doc \notin O}{\mathrm{argmax}}        
\  Score(doc) 
\label{best.doc.d}
 \end{equation}

\begin{equation} best\_sent =\underset{sent \in  best\_doc}{\mathrm{argmax}}  \  Score(sent) 
\label{best.sent.d}
 \end{equation}
 
Using documents for each stage of the process (choosing $1d$, $2d$,
and $3d$) is our baseline, and replicates BMI, except for the use of
the union of documents and sentences to compute word features.  For
brevity, we use the notation $ddd$ to represent this combination of
choices, and more generally, we use $XYZ$ to denote $1X$, $2Y$ and
$3Z$, where $X,Y,Z \in \{d,s\}$. The choices for all the eight
combinations are shown in Table~\ref{combinations.table}.

\begin{table}
\centering
\caption{Eight combinations on three binary choices.}
\label{combinations.table}
\resizebox{\textwidth}{!}{
\begin{tabular}{c|c|c|c|c}
\hline
\#No & Strategy & \begin{tabular}[c]{@{}c@{}}Present\\best\_doc\\ or best\_sent \\ to reviewer\\ ($1d$ or $1s$)\end{tabular} & \begin{tabular}[c]{@{}c@{}}Add (best\_doc, $l$)\\ or (best\_sent, $l$) \\and 100 random\\ sentences or\\documents \\ as non-relevant \\ to training set\\ ($2d$ or $2s$)\end{tabular} & \begin{tabular}[c]{@{}c@{}}Select \\ (best\_sent, best\_doc)\\ pair\\ ($3d$ or $3s$)\end{tabular} \\ \hline \hline
1 & ddd & best\_doc & \begin{tabular}[c]{@{}c@{}}$2d$: \\ (best\_doc, $l$) and \\ 100 randomly selected\\  documents treated as \\ non-relevant \end{tabular} & \begin{tabular}[c]{@{}c@{}}$3d$: \\ the highest-scoring document \\ not yet in system output, and \\ the highest-scoring sentence \\ within that document.\end{tabular} \\ \hline
2 & sdd & best\_sent & $2d$ & $3d$ \\ \hline
3 & dsd & best\_doc & \begin{tabular}[c]{@{}c@{}}$2s$: \\ (best\_sent, $l$) and\\ 100 randomly\\ selected sentences\\ treated as non-relevant \end{tabular} & $3d$ \\ \hline
4 & ssd & best\_sent & $2s$ & $3d$ \\ \hline
5 & dds & best\_doc & $2d$ & \begin{tabular}[c]{@{}c@{}}$3s$: \\ the highest-scoring sentence \\ contained in any document \\ not yet in system output, and \\ the document containing that \\ sentence.\end{tabular} \\ \hline
6 & sds & best\_sent & $2d$ & $3s$ \\ \hline
7 & dss & best\_doc & $2s$ & $3s$ \\ \hline
8 & sss & best\_sent & $2s$ & $3s$ \\ \hline
\end{tabular}
}
\end{table}

\subsection{Test Collections}
\label{section.SCAL}


\begin{table}
\centering
\caption{Dataset statistics}
\label{dataset.stat}
\begin{tabular}{c|c|r|r|r}
\hline
Dataset & \begin{tabular}[c]{@{}c@{}}Number of\\ topics\end{tabular} & \multicolumn{1}{c|}{\begin{tabular}[c]{@{}c@{}}Number of\\ documents\end{tabular}} & \multicolumn{1}{c|}{\begin{tabular}[c]{@{}c@{}}Number of\\ sentences\end{tabular}} & \multicolumn{1}{c}{\begin{tabular}[c]{@{}c@{}}Number of\\ relevant \\ documents\end{tabular}} \\ \hline
Athome1 & 10 & 290,099 & 4,616,934 & 43,980 \\ \hline
Athome2 & 10 & 460,896 & 10,493,480 & 20,005 \\ \hline
Athome3 & 10 & 902,434 & 25,622,071 & 6,429 \\ \hline
HARD & 25 & 652,309 & 10,606,819 & 1,682 \\ \hline
\end{tabular}
\end{table}

We use four test collections to evaluate the eight different
variations of continuous active learning.  We use the three 
test collections from the TREC 2015 Total Recall track: Athome1,
Athome2, and Athome3.  We also use the test collection from the 
TREC 2004 HARD track~\citep{allan2005hard,Voorhees00overviewof}.
For each collection, we used NLTK's Punkt Sentence
Tokenizer\footnote{http://www.nltk.org/api/nltk.tokenize.html} to
break all documents into sentences.  
Corpus statistics for the four collections are shown in Table
\ref{dataset.stat}.  

In order to compare sentence-level feedback with document-level
feedback strategies, we needed complete relevance labels for all
sentences as well as for all documents in the collections.  

The 2004 HARD track's collection provided pooled assessments
with complete relevance labels for all documents in
the pool.  In addition, for 25 topics, every relevant document was
divided by the TREC assessors into relevant and non-relevant passages
identified by character offsets.  For the HARD collection, we only use
the 25 topics with passage judgments.  We considered a sentence to be
relevant if it overlapped with a relevant passage.  Sentences that did
not overlap with a relevant passage were labeled non-relevant.

For both the HARD track collection and the Total Recall collections,
sentences from non-relevant and unjudged documents were labeled as
non-relevant. 

The Total Recall collections provided complete document-level
relevance judgments, i.e., the relevance of every document is known.
Each relevant document is composed of one or more relevant sentences
and zero or more non-relevant sentences.  To label the sentences as
relevant or non-relevant the first author employed ``Scalable CAL''
(``S-CAL'') \citep{cormack2016scalability} to build a calibrated
high-accuracy classifier that was used to label every sentence within
every relevant document.  Our total effort to train the S-CAL
classifier was to review 610, 453, and 376 sentences, on average, per
topic, for each of the three Athome datasets, respectively.

While neither of these methods yields a perfect labeling, their
purpose is to simulate human feedback, which is likewise imperfect.
The internal calibration of our S-CAL classifier indicated its recall
and precision both to be above $0.8$ ($F_1=0.82, 0.87, 0.81$ for
Athome1, Athome2, and Athome3, respectively), which is comparable to
human accuracy \citep{cormack2016scalability} and, we surmised, would
be good enough to test the effectiveness of sentence-level
feedback. 
Similarly, we surmised that overlap between sentences and relevant
passages in the HARD collection would yield labels that were good
enough for this purpose.

\begin{table}[]
\centering
\caption{Micro-averaged statistics of generated sentences label set on different
datasets.}
\label{label.stat}
\resizebox{\textwidth}{!}{
\begin{tabular}{c|c|c|c|c|c}
\hline
Dataset & \begin{tabular}[c]{@{}c@{}}Number of\\ sentences\\ per document\end{tabular} & \begin{tabular}[c]{@{}c@{}}Number of\\ sentences\\ per\\ relevant\\ document\end{tabular} & \begin{tabular}[c]{@{}c@{}}Number of\\ relevant\\ sentences\\ per\\ relevant\\ document\end{tabular} & \begin{tabular}[c]{@{}c@{}}Position of\\ the first\\ relevant\\ sentence\\ in relevant\\ document\end{tabular} & \begin{tabular}[c]{@{}c@{}}Proportion of\\ relevant \\ documents\\ has relevant\\ sentence\end{tabular} \\ \hline
Athome1 & 15.9 & 18.1 & 7.8 & 2.0 & 0.98 \\ \hline
Athome2 & 22.8 & 19.4 & 3.8 & 5.1 & 0.97 \\ \hline
Athome3 & 28.4 & 47.2 & 7.5 & 19.1 & 0.97 \\ \hline
HARD & 16.3 & 23.2 & 11.3 & 4.0 & 1.00 \\ \hline
\end{tabular}
}
\end{table}

\begin{figure}
\centering
\includegraphics[width=0.75\textwidth]{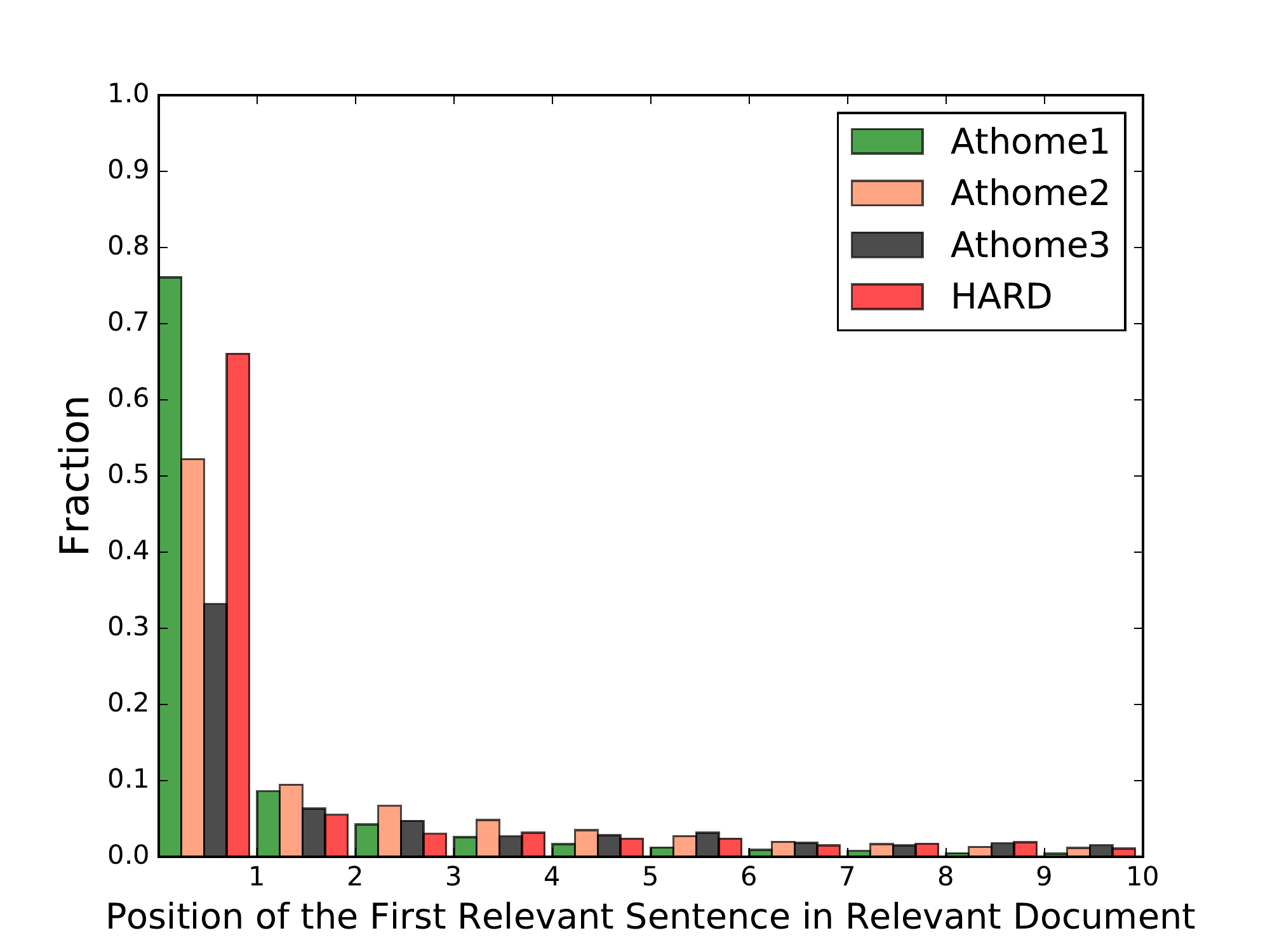}
\caption{The distribution of the position of the first relevant sentence in the 
relevant documents for different document collections.}
\label{first.rank.pos}
\end{figure}

The results of our sentence labeling are shown in
Table~\ref{label.stat}. The average position of the first relevant
sentence in each relevant document is shown in the fifth column, while
the distribution of such positions is shown in Figure
\ref{first.rank.pos}. On Athome1, Athome2 and HARD three datasets,
more than 50\% relevant documents in each dataset have their first relevant sentences located at the first sentences.
However, the position of the first relevant sentence in the relevant document is larger than $2$ for all the four datasets.
It means that the reviewer need to review more than two sentences to find the first relevant sentence in each relevant document under
the assumption that reviewer read the document sequentially.
The sixth column shows the fraction of relevant
documents containing at least one sentence labeled relevant. It shows
that nearly every relevant document contains at least one relevant sentence.


\section{Evaluation}
\label{section.eval}

The human-in-the-loop CAL simulated by the TREC Total Recall track
evaluation apparatus, which the current study adopts and extends, has
the following process.  Starting with a standard test collection
consisting of a set of documents, topic statements, and relevance
assessments (``qrels''), the most-likely relevant document is
presented to the reviewer for assessment. The reviewer's response is
simulated by consulting the qrels, and fed back to the system, which
chooses the next-most-likely-relevant document to present. The process
continues until a formal or informal stopping criterion is met,
suggesting that substantially all relevant documents have been
presented to the reviewer.

To model sentence-level feedback it was necessary to extend the
evaluation apparatus to incorporate a sentence dataset and sentence
qrels. The sentence dataset consists of all sentences extracted from
documents in the document dataset, and the sentence qrels consist of
relevance assessments for each sentence. To simulate sentence-level
feedback, the apparatus presents to the simulated reviewer a single
sentence, as determined by the system under test, and communicates the
reviewer's assessment to the system, which then selects the next
sentence for review.  The ``system-selected documents'' used for
evaluation consist of the sequence of documents from which the
sentences presented to the reviewer were extracted. In our paper,
the ``system-selected documents'' are recorded in the system output ($O$) mentioned
in the Step 6 of Algorithm~\ref{alg.gen}. The same
apparatus is used to simulate document-level feedback, except that
here, the system selects a document for presentation to the reviewer,
and the reviewer's feedback is simulated by consulting the document
qrels.  In document-level-feedback mode, the apparatus is
operationally equivalent to the TREC Total Recall apparatus.

Recall is the number of relevant documents presented to the reviewer
for assessment, as a fraction of the total number of relevant
documents ($R$), regardless of whether document- or sentence-level
feedback is employed. In our paper, the documents
presented to the reviewer are recorded by the system output ($O$).
We measure the recall at effort ($Recall@E$)
using the Equation~\ref{eq.recall}:
\begin{equation}
\label{eq.recall}
Recall@E =  \frac{|\{O@E\} \cap \textit{\{Relevant documents\}}|}{|\textit{\{Relevant documents\}}|}
\end{equation}
where the $O@E$ is the system output truncated at the effort $E$. The sets of relevant documents are the 
gold standard relevance assessments (``qrels'') provided by the TREC Total Recall 2015 Track and HARD 2004 Track
for the corresponding datasets and topics.

The Total Recall Track measured recall as a function of effort ($E$), where
effort was measured by the number of assessments rendered by the
reviewer. Gain curves were used to illustrate the overall shape of
the function, and recall at particular effort levels $a\cdot R+b$ were
tabulated, where $R$ is the number of relevant documents, $a$ is the
constant 1, 2, or 4, and $b$ is the constant 0, 100, or 1000.
Intuitively, these measures show the recall that can be achieved with
effort proportional to the number of relevant documents, plus some
fixed overhead amount.

We also measure recall as a function of effort $E$, but in this paper, we
measure effort as a linear combination of the number of assessments
rendered by the reviewer $E_{judge}$, and the number of sentences that must be
read by the reviewer to render a judgment $E_{sent}$.  If a simulated reviewer
provides an assessment on a single sentence, the reviewer reads one
sentence and makes one assessment.  When a full document is presented
for assessment, we simulate the reviewer to read the document
sequentially from the beginning to the first relevant sentence and
then make one assessment.  If the document is non-relevant, the
assessor needs to read all of the sentences in the document.

The ratio of effort required to make an assessment 
to the effort required to read a sentence
is not necessarily $1.0$.  To explore different 
ratios of effort, we express effort, $E_\lambda$, as a linear combination
of $E_{judge}$ and $E_{sent}$:
\begin{equation}
\label{eq.lambda}
E_\lambda = (1-\lambda)\cdot E_{judge} + \lambda\cdot E_{sent}
\end{equation}
where $E_{judge}$ is the number of assessments and $E_{sent}$ is the
number of sentences read.  At one extreme, we only care about the
number of assessments, i.e., $E_0 = E_{judge}$.  At the other
extreme, we only count reading effort, i.e., $E_1 = E_{sent}$.
For sentence-level feedback, $E_{judge} = E_{sent} = E_\lambda$,
regardless of $\lambda$. For document-level feedback, $E_{judge} \le
E_\lambda \le E_{sent}$, and $E_{\lambda_1} \le E_{\lambda_2}$ where
$\lambda_1 \le \lambda_2$.

For single assessment on each document $d$, the number of assessments on $d$ is $E_{judge} = 1$. 
We can simplify the assessment effort defined in Equation~\ref{eq.lambda} for a single document $d$ as
$E_\lambda =  1+ \lambda \cdot (E_{sent} - 1)$. If the $E_{sent} > 1$ for the document $d$, then
$E_\lambda > 1$. With the number of sentences needed $E_{sent}$ for reviewing this document $d$ increasing, 
the $E_\lambda$ also increases.

\section{Results}
\label{section.results}

\begin{figure}
\centering
\includegraphics[width=0.75\textwidth]{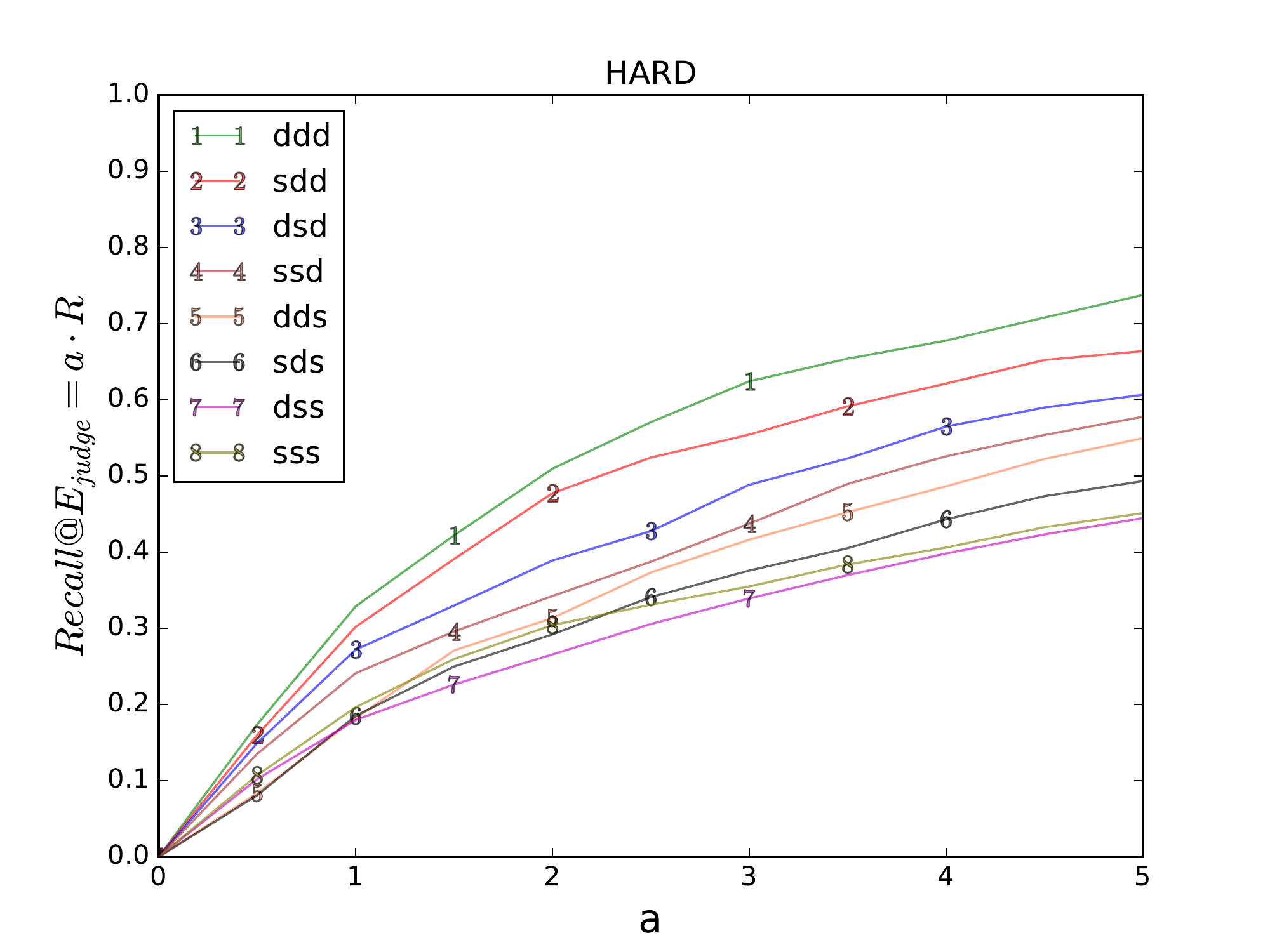}
\caption{Recall at $E_{judge}=a\cdot R$ for increasing $a$ on HARD.}
\label{AR.Doc.HARDs}
\end{figure}

\begin{figure}
\centering
\includegraphics[width=0.75\textwidth]{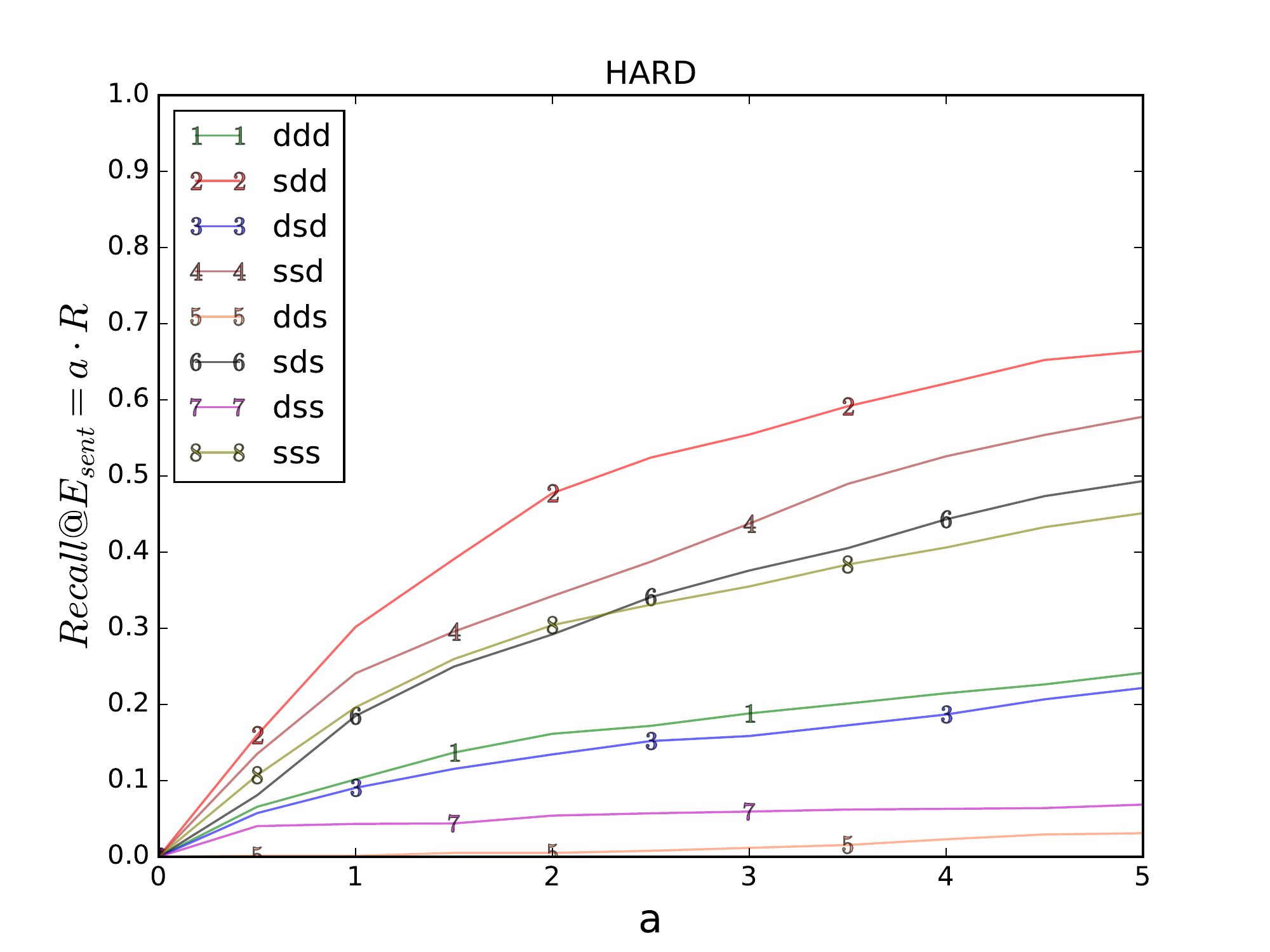}
\caption{Recall at $E_{sent}=a\cdot R$ for increasing $a$ on HARD.}
\label{AR.Sent.HARDs}
\end{figure}

We compared the sentence-level feedback strategies
with the document-level feedback strategies on three different
dimensions---in total, eight combinations shown in Table~\ref{combinations.table}.  As explained in
Section~\ref{section.eval}, we measure performance as recall versus
effort.  At one extreme, we can measure effort as the number of
assessments (judgments) made by the reviewer, i.e., effort =
$E_{judge}$.  At the other extreme, we can measure effort as the
number of sentences read, i.e., effort = $E_{sent}$.  

\begin{figure}[h!]
\minipage{0.49\textwidth}
\includegraphics[width=\linewidth]{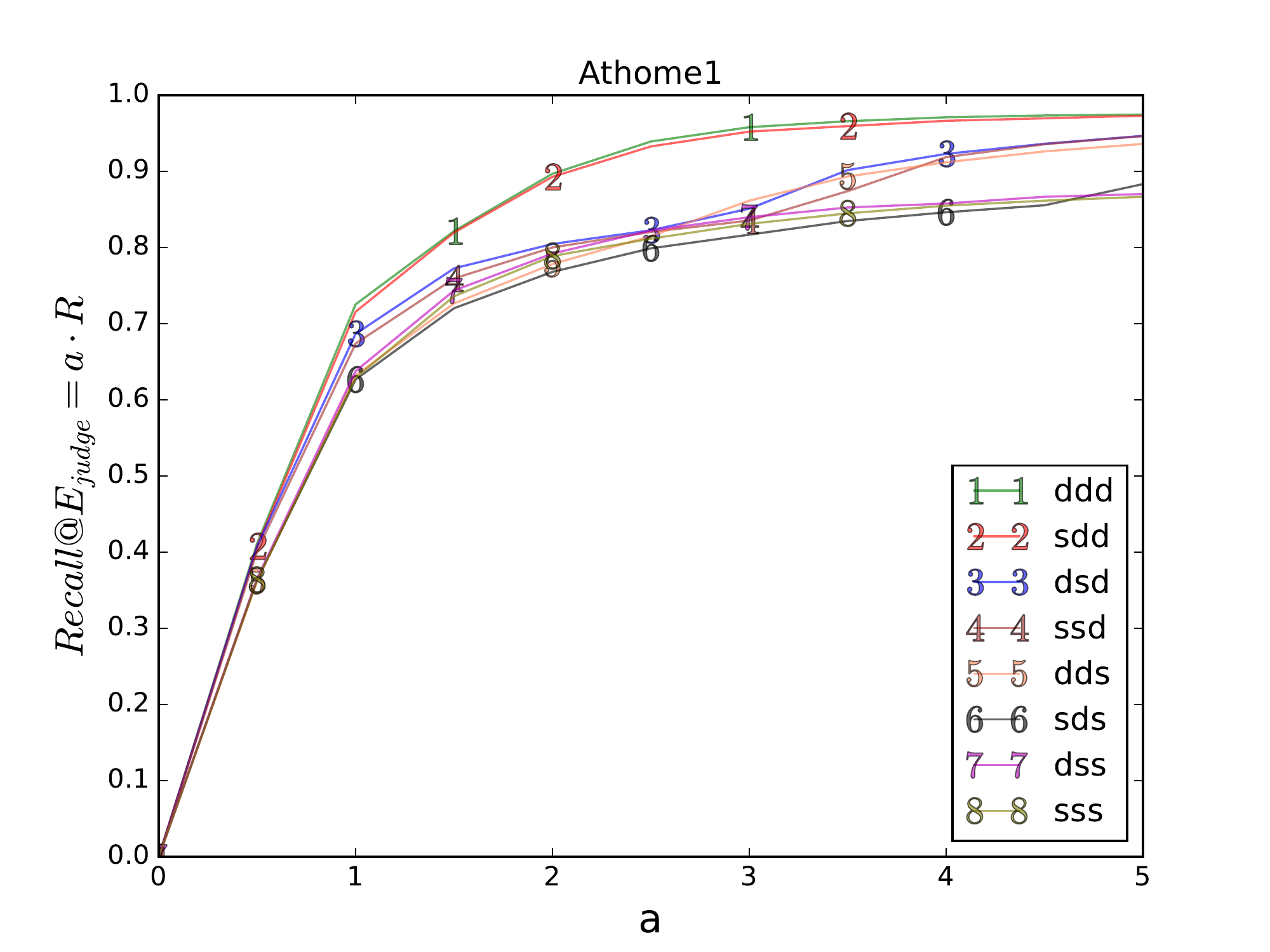}
\caption{Recall at $E_{judge}=a\cdot R$ with varying $a$ on Athome1.}
\label{AR.Doc.a1}
\endminipage\hfill
\minipage{0.49\textwidth}
\includegraphics[width=\linewidth]{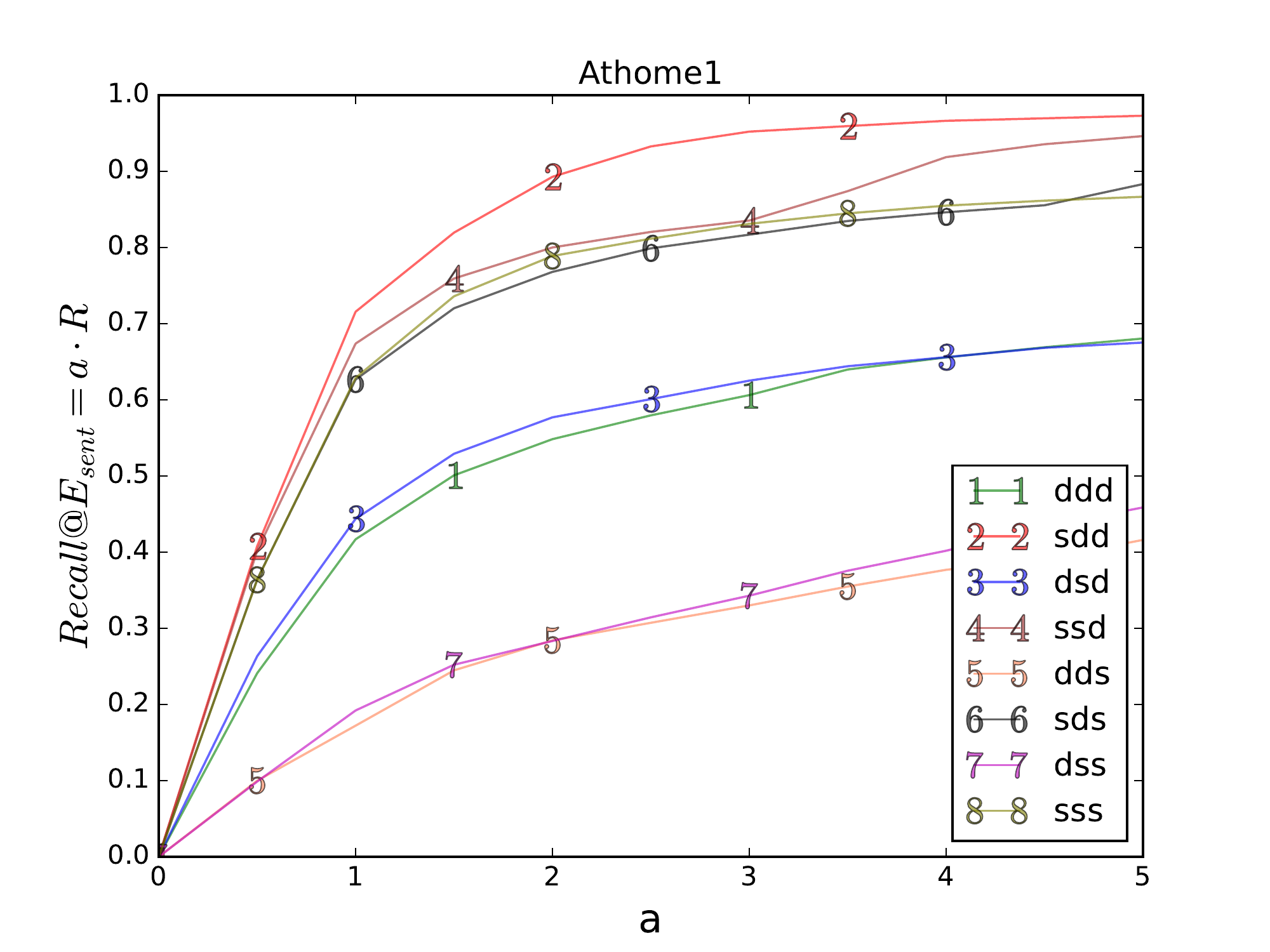}
\caption{Recall at $E_{sent}=a\cdot R$ with varying $a$ on Athome1.}
\label{AR.Sent.a1}
\endminipage\hfill
\minipage{0.49\textwidth}
\includegraphics[width=\linewidth]{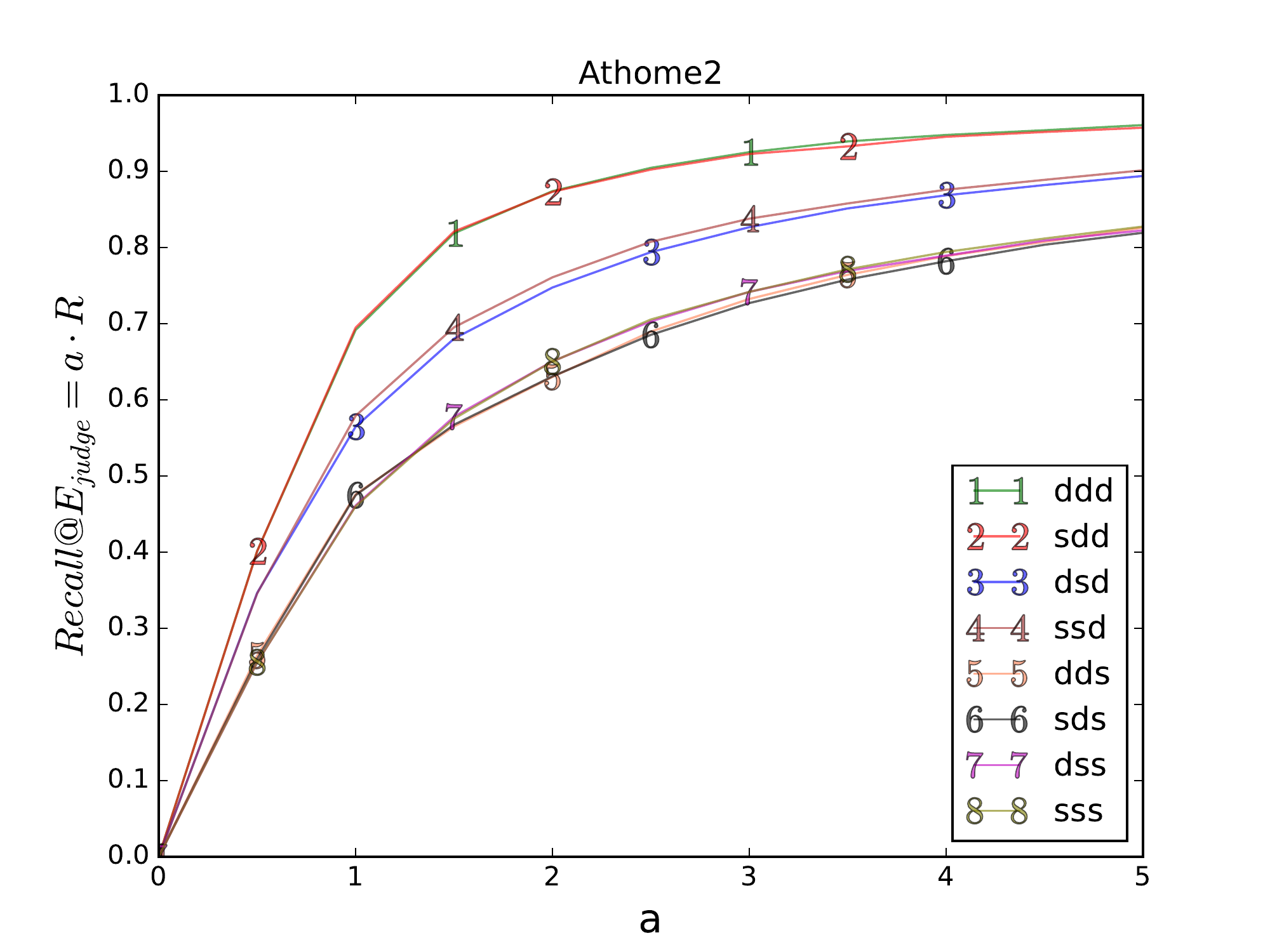}
\caption{Recall at $E_{judge}=a\cdot R$ with varying $a$ on Athome2.}
\label{AR.Doc.a2}
\endminipage\hfill
\minipage{0.49\textwidth}
\includegraphics[width=\linewidth]{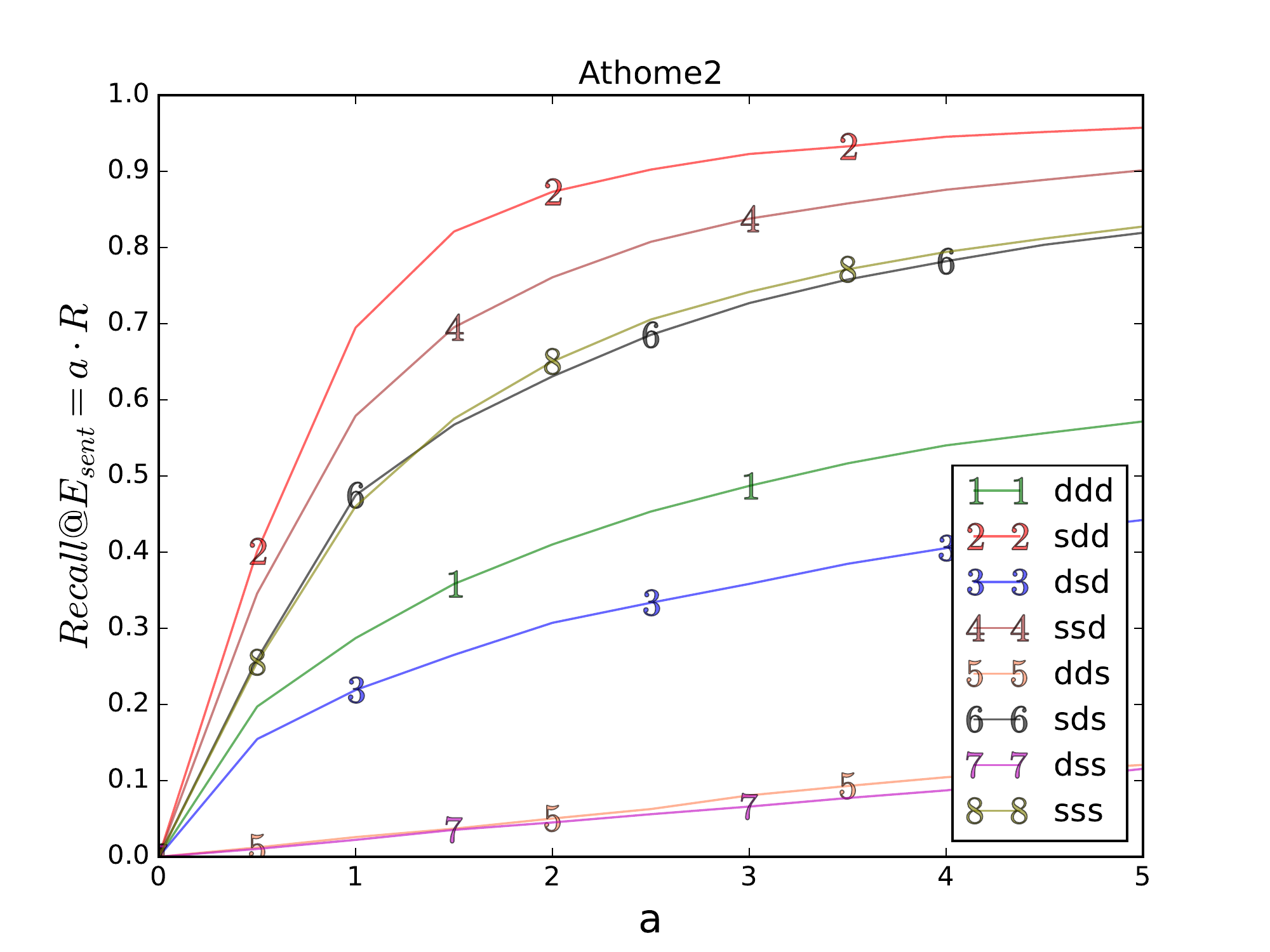}
\caption{Recall at $E_{sent}=a\cdot R$ with varying $a$ on Athome2.}
\label{AR.Sent.a2}
\endminipage\hfill
\minipage{0.49\textwidth}
\includegraphics[width=\linewidth]{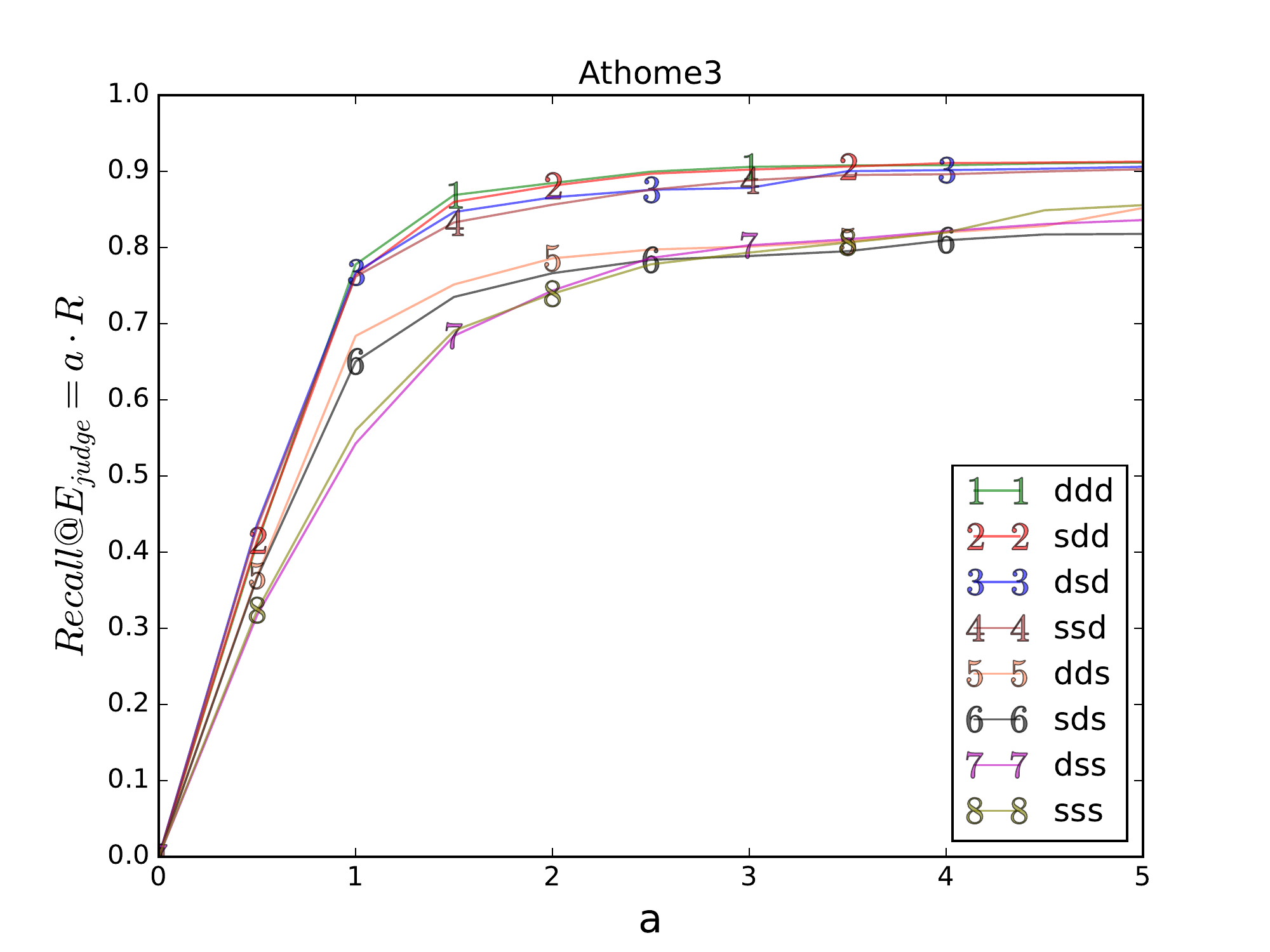}
\caption{Recall at $E_{judge}=a\cdot R$ with varying $a$ on Athome3.}
\label{AR.Doc.a3}
\endminipage\hfill
\minipage{0.49\textwidth}
\includegraphics[width=\linewidth]{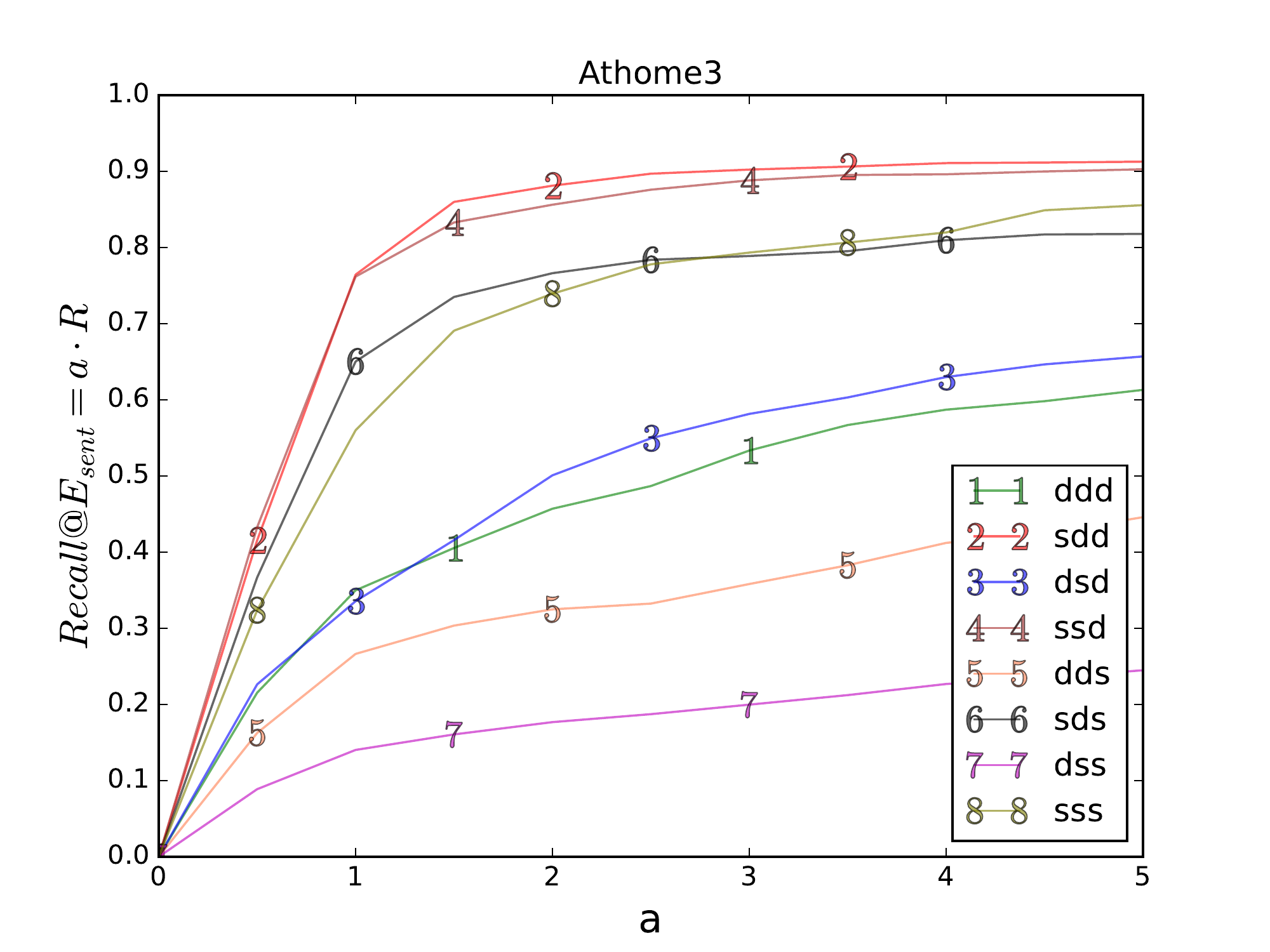}
\caption{Recall at $E_{sent}=a\cdot R$ with varying $a$ on Athome3.}
\label{AR.Sent.a3}
\endminipage\hfill
\end{figure}

Figures~\ref{AR.Doc.HARDs} and~\ref{AR.Sent.HARDs} show recall vs.\ effort for the HARD test
collection.  Figure~\ref{AR.Doc.HARDs} measures effort as a function
of the number of judgments ($E_{judge}$), where the horizontal axis
reports judgments in multiples of the number of relevant document
$R$.  For example, $aR$ documents, where $a = 2$ means that twice as
many judgments have been made as there are relevant documents.
Figure~\ref{AR.Sent.HARDs} measures effort as a function of the number
of sentences read ($E_{sent}$).  The equivalent plots for the Athome
collections are found at the end of the paper in Figures
\ref{AR.Doc.a1} --~\ref{AR.Sent.a3}.

In general, when effort is measured in terms of judgments only ($E\_{judge}$), 
we find that the training on and selecting documents to be superior to other methods
regardless whether the reviewer judged documents ($ddd$ strategy) or sentences ($sdd$ strategy),
across all eight combinations, for all four datasets, for all $a$.  We also
find that training on sentences with the selection of documents ($dsd$ and $ssd$)
strategies to be worse than the strategies that training on documents and selecting
documents ($ddd$ and $sdd$) on all
datasets, but superior to the other four strategies: $dds$, $dss$, $sds$, and
$sss$. The overall comparison of judgment effort for all the eight combinations is that
$\{ddd, sdd\} > \{dsd, ssd\} > \{dds, sds, dss, sss\}$.

When effort is measured in terms of sentences read only ($E\_{sent}$), all of the
sentence-level feedback strategies in which reviewer judges documents $\{sdd,ssd,sds,sss\}$ achieve much higher recall than the
document-level feedback strategies in which reviewer judges sentences $\{ddd,dsd,dds,dss\}$ for a given level of effort, as
measured in terms of the number of sentences reviewed. Among the four
sentence-level feedback strategies, $sdd$ is superior, and the relative
effectiveness among the sentence-based strategies is consistent with
the result when effort is measured by the number of assessments. The overall ranking
of four sentence-level feedback strategies evaluated by number of sentences read is $\{sdd\} > \{ssd\} > \{sds, sss\}$.

These results suggest that training using documents and selecting
the highest-ranking document from the document-rank list to review  ($ddd$ and $sdd$)
will lead to superior results over other strategies, regardless of
whether sentences or documents are presented to the reviewer for
feedback. At the same time, the choice of using sentences ($sdd$) or documents ($ddd$)
for feedback has very little impact on the recall that can be achieved
for a given number of assessments.

\begin{table}
\centering
\caption{Recall at $E_{judge}=R$, $E_{0.5}=R$, and $E_{sent}=R$ for different strategies on different datasets. 
We bold the greater value if the difference in recall between $sdd$ and $ddd$ is statistically significant. The overall
is the average result over all the 55 topics from all the four datasets.}
\label{1r-recall-doc-table}
\begin{tabular}{c|c|c|c|c|c|c|c|c|c}
\hline
Dataset                  & Effort    & ddd  & sdd           & dsd  & ssd  & dds  & sds  & dss  & sss  \\ \hline \hline
\multirow{3}{*}{Athome1} & 1R\_Judge & 0.73 & 0.72          & 0.69 & 0.67 & 0.63 & 0.63 & 0.64 & 0.63 \\  
                         & 1R\_0.5   & 0.48 & \textbf{0.72} & 0.51 & 0.67 & 0.24 & 0.63 & 0.26 & 0.63 \\  
                         & 1R\_Sent  & 0.42 & \textbf{0.72} & 0.44 & 0.67 & 0.17 & 0.63 & 0.19 & 0.63 \\ \hline
\multirow{3}{*}{Athome2} & 1R\_Judge & 0.69 & 0.69          & 0.56 & 0.58 & 0.48 & 0.47 & 0.46 & 0.46 \\  
                         & 1R\_0.5   & 0.36 & \textbf{0.69} & 0.27 & 0.58 & 0.05 & 0.47 & 0.04 & 0.46 \\  
                         & 1R\_Sent  & 0.29 & \textbf{0.69} & 0.22 & 0.58 & 0.03 & 0.47 & 0.02 & 0.46 \\ \hline
\multirow{3}{*}{Athome3} & 1R\_Judge & 0.78 & 0.76          & 0.77 & 0.76 & 0.68 & 0.65 & 0.54 & 0.56 \\  
                         & 1R\_0.5   & 0.40 & \textbf{0.76} & 0.41 & 0.76 & 0.30 & 0.65 & 0.16 & 0.56 \\  
                         & 1R\_Sent  & 0.35 & \textbf{0.76} & 0.34 & 0.76 & 0.27 & 0.65 & 0.14 & 0.56 \\ \hline
\multirow{3}{*}{HARD}    & 1R\_Judge & 0.34 & 0.31          & 0.28 & 0.25 & 0.19 & 0.19 & 0.19 & 0.20 \\  
                         & 1R\_0.5   & 0.15 & \textbf{0.31} & 0.13 & 0.25 & 0.01 & 0.19 & 0.05 & 0.20 \\  
                         & 1R\_Sent  & 0.11 & \textbf{0.31} & 0.09 & 0.25 & 0.00 & 0.19 & 0.04 & 0.20 \\ \hline \hline
\multirow{3}{*}{Overall} & 1R\_Judge & 0.55 & 0.54          & 0.50 & 0.48 & 0.41 & 0.41 & 0.38 & 0.39 \\  
                         & 1R\_0.5   & 0.29 & \textbf{0.54} & 0.28 & 0.48 & 0.11 & 0.41 & 0.11 & 0.39 \\  
                         & 1R\_Sent  & 0.24 & \textbf{0.54} & 0.22 & 0.48 & 0.09 & 0.41 & 0.08 & 0.39 \\ \hline
\end{tabular}
\end{table}

\begin{table}
\centering
\caption{Recall at $E_{judge}=2R$, $E_{0.5}=2R$, and $E_{sent}=2R$}
\label{2r-recall-doc-table}
\begin{tabular}{c|c|c|c|c|c|c|c|c|c}
\hline
Dataset                  & Effort    & ddd  & sdd           & dsd  & ssd  & dds  & sds  & dss  & sss  \\ \hline \hline
\multirow{3}{*}{Athome1} & 2R\_Judge & 0.90 & 0.89          & 0.80 & 0.80 & 0.78 & 0.77 & 0.79 & 0.79 \\  
                         & 2R\_0.5   & 0.63 & \textbf{0.89} & 0.63 & 0.80 & 0.36 & 0.77 & 0.37 & 0.79 \\  
                         & 2R\_Sent  & 0.55 & \textbf{0.89} & 0.58 & 0.80 & 0.28 & 0.77 & 0.28 & 0.79 \\ \hline
\multirow{3}{*}{Athome2} & 2R\_Judge & 0.87 & 0.87          & 0.75 & 0.76 & 0.63 & 0.63 & 0.65 & 0.65 \\  
                         & 2R\_0.5   & 0.51 & \textbf{0.87} & 0.38 & 0.76 & 0.10 & 0.63 & 0.08 & 0.65 \\  
                         & 2R\_Sent  & 0.41 & \textbf{0.87} & 0.31 & 0.76 & 0.05 & 0.63 & 0.05 & 0.65 \\ \hline
\multirow{3}{*}{Athome3} & 2R\_Judge & 0.88 & 0.88          & 0.87 & 0.86 & 0.79 & 0.77 & 0.74 & 0.74 \\  
                         & 2R\_0.5   & 0.55 & \textbf{0.88} & 0.59 & 0.86 & 0.39 & 0.77 & 0.22 & 0.74 \\  
                         & 2R\_Sent  & 0.46 & \textbf{0.88} & 0.50 & 0.86 & 0.32 & 0.77 & 0.18 & 0.74 \\ \hline
\multirow{3}{*}{HARD}    & 2R\_Judge & 0.53 & 0.50          & 0.41 & 0.36 & 0.33 & 0.30 & 0.28 & 0.32 \\  
                         & 2R\_0.5   & 0.21 & \textbf{0.50} & 0.18 & 0.36 & 0.02 & 0.30 & 0.07 & 0.32 \\  
                         & 2R\_Sent  & 0.17 & \textbf{0.50} & 0.14 & 0.36 & 0.01 & 0.30 & 0.06 & 0.32 \\ \hline \hline
\multirow{3}{*}{Overall} & 2R\_Judge & 0.72 & 0.71          & 0.62 & 0.60 & 0.55 & 0.53 & 0.52 & 0.54 \\  
                         & 2R\_0.5   & 0.40 & \textbf{0.71} & 0.37 & 0.60 & 0.16 & 0.53 & 0.15 & 0.54 \\  
                         & 2R\_Sent  & 0.33 & \textbf{0.71} & 0.32 & 0.60 & 0.12 & 0.53 & 0.12 & 0.54 \\ \hline
\end{tabular}
\end{table}

\begin{table}
\centering
\caption{Recall at $E_{judge}=4R$, $E_{0.5}=4R$, and $E_{sent}=4R$}
\label{4r-recall-doc-table}
\begin{tabular}{c|c|c|c|c|c|c|c|c|c}
\hline 
Dataset                  & Effort    & ddd  & sdd           & dsd  & ssd  & dds  & sds  & dss  & sss  \\ \hline \hline
\multirow{3}{*}{Athome1} & 4R\_Judge & 0.97 & 0.97          & 0.92 & 0.92 & 0.91 & 0.85 & 0.86 & 0.85 \\  
                         & 4R\_0.5   & 0.72 & \textbf{0.97} & 0.71 & 0.92 & 0.49 & 0.85 & 0.55 & 0.85 \\  
                         & 4R\_Sent  & 0.66 & \textbf{0.97} & 0.66 & 0.92 & 0.38 & 0.85 & 0.40 & 0.85 \\ \hline
\multirow{3}{*}{Athome2} & 4R\_Judge & 0.95 & 0.95          & 0.87 & 0.88 & 0.79 & 0.78 & 0.79 & 0.79 \\  
                         & 4R\_0.5   & 0.64 & \textbf{0.95} & 0.51 & 0.88 & 0.18 & 0.78 & 0.18 & 0.79 \\  
                         & 4R\_Sent  & 0.54 & \textbf{0.95} & 0.41 & 0.88 & 0.10 & 0.78 & 0.09 & 0.79 \\ \hline
\multirow{3}{*}{Athome3} & 4R\_Judge & 0.91 & 0.91          & 0.90 & 0.90 & 0.82 & 0.81 & 0.82 & 0.82 \\  
                         & 4R\_0.5   & 0.67 & \textbf{0.91} & 0.72 & 0.90 & 0.50 & 0.81 & 0.30 & 0.82 \\  
                         & 4R\_Sent  & 0.59 & \textbf{0.91} & 0.63 & 0.90 & 0.41 & 0.81 & 0.23 & 0.82 \\ \hline
\multirow{3}{*}{HARD}    & 4R\_Judge & 0.71 & 0.65          & 0.59 & 0.55 & 0.51 & 0.46 & 0.41 & 0.42 \\  
                         & 4R\_0.5   & 0.29 & \textbf{0.65} & 0.26 & 0.55 & 0.05 & 0.46 & 0.08 & 0.42 \\  
                         & 4R\_Sent  & 0.22 & \textbf{0.65} & 0.19 & 0.55 & 0.02 & 0.46 & 0.07 & 0.42 \\ \hline \hline
\multirow{3}{*}{Overall} & 4R\_Judge & 0.83 & 0.81          & 0.76 & 0.74 & 0.69 & 0.65 & 0.64 & 0.64 \\  
                         & 4R\_0.5   & 0.50 & \textbf{0.81} & 0.47 & 0.74 & 0.24 & 0.65 & 0.22 & 0.64 \\  
                         & 4R\_Sent  & 0.43 & \textbf{0.81} & 0.40 & 0.74 & 0.17 & 0.65 & 0.16 & 0.64 \\ \hline
\end{tabular}
\end{table}

\begin{table}
\centering
\caption{recall[$sdd$]-recall[$ddd$] at effort = $a\cdot E_{judge}$ ($95\%$ Confidence interval).}
\label{ddd-sdd-ci-table}
\begin{tabular}{c|c|c|c}
\hline
Dataset & a=1             & a=2             & a=4             \\ \hline \hline
Athome1 & (-0.025, 0.006) & (-0.012, 0.003) & (-0.009, -0.0003)\tablefootnote{The mean difference between recall[$sdd$] and recall[ddd] equals 0.0046 and $p=0.037$ at effort = $4\cdot E_{judge}$ on Athome1.}  \\ \hline
Athome2 & (-0.008, 0.014) & (-0.005, 0.003) & (-0.007, 0.002) \\ \hline
Athome3 & (-0.043, 0.016) & (-0.015, 0.008) & (-0.005, 0.011) \\ \hline
HARD    & (-0.074, 0.020)  & (-0.071, 0.007) & (-0.122, 0.009) \\ \hline \hline
Overall & (-0.037, 0.006) & (-0.034, 0.002) & (-0.056, 0.003) \\ \hline 
\end{tabular}
\end{table}

\begin{table}
\centering
\caption{recall[$sdd$]-recall[$ddd$] at effort = $a\cdot E_{sent}$ ($95\%$ Confidence interval).}
\label{ddd-sdd--sent-ci-table}
\begin{tabular}{c|c|c|c}
\hline
Dataset & a=1 & a=2 & a=4 \\ \hline \hline
Athome1 & (0.178, 0.420) & (0.181, 0.508) & (0.107, 0.514) \\ \hline
Athome2 & (0.308, 0.508) & (0.352, 0.574) & (0.266, 0.545) \\ \hline
Athome3 & (0.292, 0.537) & (0.244, 0.605) & (0.148, 0.499) \\ \hline
HARD & (0.121, 0.279) & (0.222, 0.410) & (0.297, 0.516) \\ \hline \hline
Overall & (0.242, 0.348) & (0.307, 0.428) & (0.306, 0.442) \\ \hline
\end{tabular}
\end{table}

\begin{table}
\centering
\caption{recall[$sdd$]-recall[$ddd$] at effort = $a\cdot E_{0.5}$ ($95\%$ Confidence interval).}
\label{ddd-sdd--lambda-ci-table}
\begin{tabular}{c|c|c|c}
\hline
Dataset & a=1            & a=2            & a=4            \\ \hline \hline
Athome1 & (0.121, 0.344) & (0.127, 0.393) & (0.041, 0.445) \\ \hline
Athome2 & (0.210, 0.360) & (0.227, 0.401) & (0.163, 0.390)  \\ \hline
Athome3 & (0.250, 0.373) & (0.246, 0.394) & (0.178, 0.349) \\ \hline
HARD    & (0.092, 0.225) & (0.193, 0.365) & (0.249, 0.445) \\ \hline \hline
Overall & (0.194, 0.290) & (0.247, 0.356) & (0.238, 0.365) \\ \hline
\end{tabular}
\end{table}

\begin{figure*}
\centering
\includegraphics[width=0.75\textwidth]{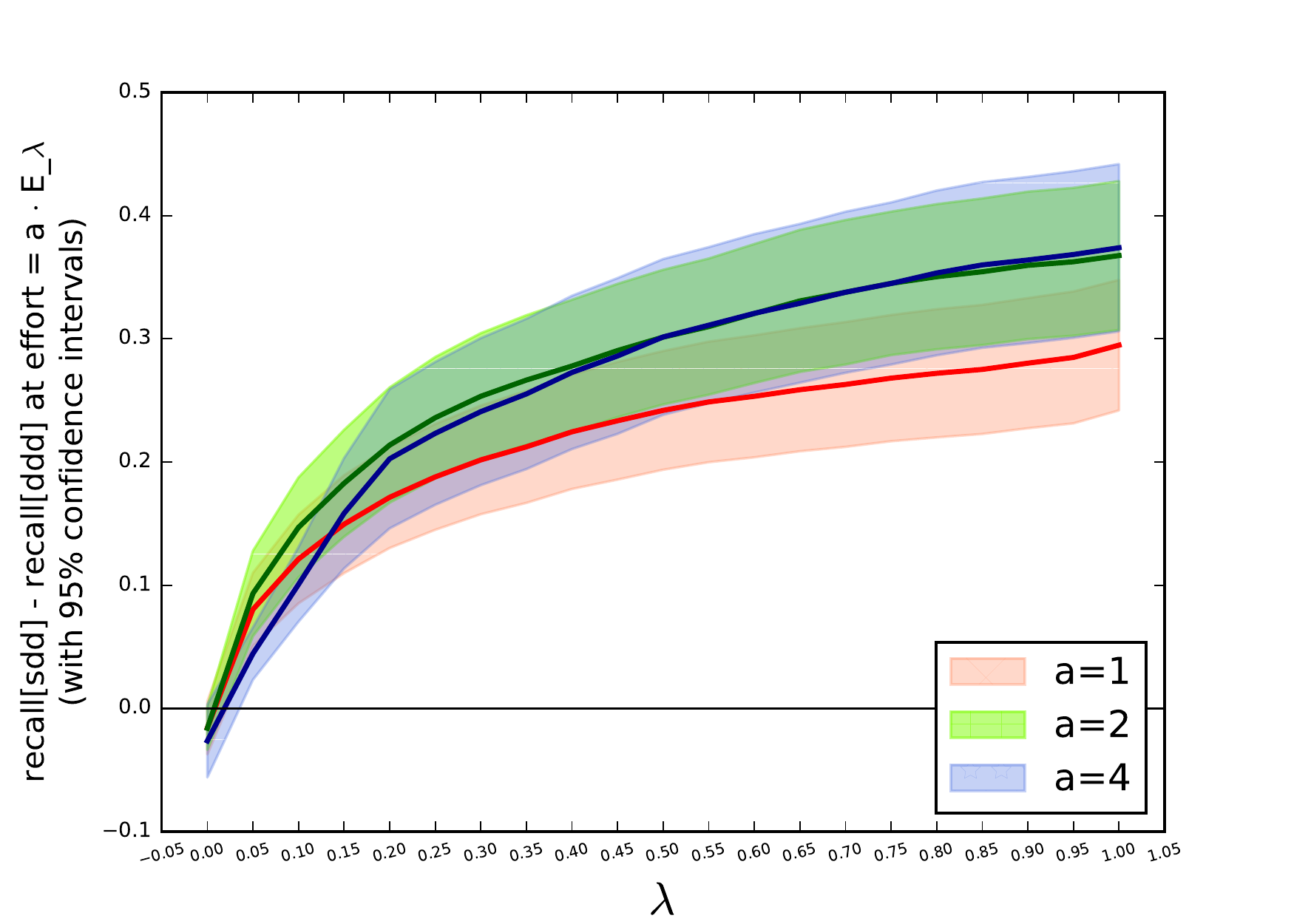}
\caption{recall[$sdd$]- recall[$ddd$] at $E_\lambda=aR, \text{where } a\in\{1,2,4\}$ by varying
$\lambda$ from 0 to 1 by step size 0.05 ($95\%$ Confidence interval). $E_\lambda=E_{judge}$ where $\lambda = 0$ and $E_\lambda=E_{sent}$ where $\lambda = 1$. With the increase of $\lambda$, recall[$sdd$] became 
significantly larger than recall[$ddd$] for all values of $a$.}
\label{ddd.sdd.ci}
\end{figure*}

The actual recall achieved by each method at multiples of $R$ is
reported in Table~\ref{1r-recall-doc-table} ($1R$),
Table~\ref{2r-recall-doc-table} ($2R$), and Table~\ref{4r-recall-doc-table}
($4R$).  These tables also report effort as a equal combination of
number of judgments and number of sentences read ($recall@E_\lambda$,
where $\lambda=0.5$).  In each table, we compare the $ddd$ and $sdd$
methods and if the difference in recall is statistically
significant, we bold the greater value.  We measure statistical
significance with a two-sided, Student's t-test and significance is
for p-values less than 0.05.  For example, in
Table~\ref{1r-recall-doc-table}, when effort is equal to the number of
relevant documents ($1R$) and measured by the number of sentences read
(1R\_Sent) on Athome1, the $ddd$ (recall=0.42) and $sdd$
(recall=0.72) methods are different at a statistically significant
level.   

The most interesting observation to be made from
Tables~\ref{1r-recall-doc-table},~\ref{2r-recall-doc-table}, and
\ref{4r-recall-doc-table} is that when effort is measured in number of
judgments $E_{judge}$, $sdd$ and $ddd$ are usually equivalent, and when effort is
measured in number of sentences read $E_{sent}$, $sdd$ is vastly superior to $ddd$.
What this means is that for essentially the same number of judgments,
we can achieve the same level of recall by only judging the best
sentence from a document --- we do not have to bother examining the
entire document to judge its relevance.

Defined in Equation~\ref{eq.lambda}, effort $E_\lambda$ is a function of the number of assessments $E_{judge}$
and the number of sentences read $E_{sent}$. 
We also calculate the $95\%$ confidence
interval for the difference of $recall@E_{0.5}=a\cdot
R$ between $ddd$ and $sdd$.  We find that $recall@E_\lambda=a\cdot R$
is significantly better for $sdd$ than $ddd$ for all values of $a$
when $\lambda =0.5$. We show the confidence interval of the difference
between $ddd$ and $sdd$ for different effort measurements $E_{judge}$, $E_{sent}$ and $E_\lambda$ with 
various values of $a$ in Tables
\ref{ddd-sdd-ci-table},~\ref{ddd-sdd--sent-ci-table}, and
\ref{ddd-sdd--lambda-ci-table}.

To get a better sense of when $sdd$ becomes superior to $ddd$, we
varied $\lambda$ from $0$ to $1$ by step size $0.05$ and plotted in Figure~\ref{ddd.sdd.ci} the 
$95\%$ confidence interval for the difference of
$recall@E_\lambda=a\cdot R$ between $ddd$ and $sdd$. As can be seen, once
the cost of reading sentences starts to have some weight where $\lambda=0.05$, $sdd$ becomes
superior to $ddd$. The recall[$sdd$]- recall[$ddd$] became larger with the increase of 
$\lambda$.

For single assessment on each document $d$, we can simplify the effort $E_{\lambda=0.05}$ for document $d$ as
$E_{\lambda=0.05} =  1+ 0.05 \cdot (E_{sent} - 1)$. 
As shown in Table~\ref{label.stat}, the position of the first relevant sentence in the relevant document
is always greater than $2.0$. Based our assumption that the reviewer read the document sequentially from the 
beginning to the first relevant sentence, we can infer that $E_{sent} \geq 2.0$.
To make this more concrete, if the number of sentences reviewed $E_{sent}$ 
for $d$ is more than $1$, $sdd$ can be superior than $ddd$ in terms of effort to achieve the same level of recall. 
In other words, if the time to judge a document is substantively more than judging a sentence, $sdd$ could be
more efficient than $ddd$.

\section{Conclusions}
\label{section.conclusion}
This simulation study suggests that an active learning method can identify
a single sentence from each document that contains sufficient information
for a user to assess the relevance of the document for the purpose of
relevance feedback.  The best-performing active learning method selected
for assessment the highest-scoring sentence from the highest-scoring document,
based on a model trained using entire documents whose labels were determined
exclusively from a single sentence.

Under the assumption that the user can review a sentence more quickly than
an entire document, the results of our study suggest that a system in which
only sentences were presented to the user would achieve very high recall more
quickly than a system in which entire documents were presented.

The synthetic labels used to simulate user feedback were imperfect, but of comparable
quality, according to recall and precision, to what has been observed for human 
users \citep{voorhees2000variations}.



\begin{thebibliography}{60}
\providecommand{\natexlab}[1]{#1}
\providecommand{\url}[1]{{#1}}
\providecommand{\urlprefix}{URL }
\expandafter\ifx\csname urlstyle\endcsname\relax
  \providecommand{\doi}[1]{DOI~\discretionary{}{}{}#1}\else
  \providecommand{\doi}{DOI~\discretionary{}{}{}\begingroup
  \urlstyle{rm}\Url}\fi
\providecommand{\eprint}[2][]{\url{#2}}

\bibitem[{Aalbersberg(1992)}]{aalbersberg1992incremental}
Aalbersberg IJ (1992) Incremental relevance feedback. In: Proceedings of the
  15th annual international ACM SIGIR conference on Research and development in
  information retrieval, ACM, pp 11--22

\bibitem[{Abualsaud et~al(2018{\natexlab{a}})Abualsaud, Cormack, Ghelani,
  Ghenai, Grossman, Rahbariasl, Smucker, and Zhang}]{zhang2018trec}
Abualsaud M, Cormack GV, Ghelani N, Ghenai A, Grossman MR, Rahbariasl S,
  Smucker MD, Zhang H (2018{\natexlab{a}}) Uwaterloomds at the trec 2018 common
  core track. TREC

\bibitem[{Abualsaud et~al(2018{\natexlab{b}})Abualsaud, Ghelani, Zhang,
  Smucker, Cormack, and Grossman}]{DBLP:conf/sigir/AbualsaudGZSCG18}
Abualsaud M, Ghelani N, Zhang H, Smucker MD, Cormack GV, Grossman MR
  (2018{\natexlab{b}}) A system for efficient high-recall retrieval. In: The
  41st International {ACM} {SIGIR} Conference on Research {\&} Development in
  Information Retrieval, {SIGIR} 2018, Ann Arbor, MI, USA, July 08-12, 2018, pp
  1317--1320, \doi{10.1145/3209978.3210176},
  \urlprefix\url{http://doi.acm.org/10.1145/3209978.3210176}

\bibitem[{Allan(2005)}]{allan2005hard}
Allan J (2005) Hard track overview in trec 2003 high accuracy retrieval from
  documents. Tech. rep., DTIC Document

\bibitem[{Baron et~al(2006)Baron, Lewis, and Oard}]{baron2006trec}
Baron JR, Lewis DD, Oard DW (2006) Trec 2006 legal track overview. In: TREC

\bibitem[{Baruah et~al(2016)Baruah, Zhang, Guttikonda, Lin, Smucker, and
  Vechtomova}]{baruah2016optimizing}
Baruah G, Zhang H, Guttikonda R, Lin J, Smucker MD, Vechtomova O (2016)
  Optimizing nugget annotations with active learning. In: Proceedings of the
  25th ACM International on Conference on Information and Knowledge Management,
  ACM, pp 2359--2364

\bibitem[{Blair and Maron(1985)}]{blair1985evaluation}
Blair DC, Maron ME (1985) An evaluation of retrieval effectiveness for a
  full-text document-retrieval system. Communications of the ACM 28(3):289--299

\bibitem[{B{\"u}ttcher et~al(2006)B{\"u}ttcher, Clarke, and
  Soboroff}]{buttcher2006trec}
B{\"u}ttcher S, Clarke CL, Soboroff I (2006) The trec 2006 terabyte track. In:
  TREC, vol~6, p~39

\bibitem[{B\"{u}ttcher et~al(2007)B\"{u}ttcher, Clarke, Yeung, and
  Soboroff}]{Buttcher:2007:RIR:1277741.1277755}
B\"{u}ttcher S, Clarke CLA, Yeung PCK, Soboroff I (2007) Reliable information
  retrieval evaluation with incomplete and biased judgements. In: Proceedings
  of the 30th Annual International ACM SIGIR Conference on Research and
  Development in Information Retrieval, ACM, New York, NY, USA, SIGIR '07, pp
  63--70, \doi{10.1145/1277741.1277755},
  \urlprefix\url{http://doi.acm.org/10.1145/1277741.1277755}

\bibitem[{Clarke et~al(2005)Clarke, Scholer, and Soboroff}]{clarke2005trec}
Clarke CL, Scholer F, Soboroff I (2005) The trec 2005 terabyte track. In: TREC

\bibitem[{Cormack and Grossman(2014)}]{cormack2014evaluation}
Cormack GV, Grossman MR (2014) Evaluation of machine-learning protocols for
  technology-assisted review in electronic discovery. In: Proceedings of the
  37th international ACM SIGIR conference on Research \& development in
  information retrieval, ACM, pp 153--162

\bibitem[{Cormack and Grossman(2015)}]{DBLP:journals/corr/CormackG15}
Cormack GV, Grossman MR (2015) Autonomy and reliability of continuous active
  learning for technology-assisted review. CoRR abs/1504.06868,
  \urlprefix\url{http://arxiv.org/abs/1504.06868}

\bibitem[{Cormack and Grossman(2016)}]{cormack2016scalability}
Cormack GV, Grossman MR (2016) Scalability of continuous active learning for
  reliable high-recall text classification. In: Proceedings of the 25th ACM
  International on Conference on Information and Knowledge Management, ACM, pp
  1039--1048

\bibitem[{Cormack and Grossman(2017{\natexlab{a}})}]{cormack2017navigating}
Cormack GV, Grossman MR (2017{\natexlab{a}}) Navigating imprecision in
  relevance assessments on the road to total recall: Roger and me. In:
  Proceedings of the 40th International ACM SIGIR Conference on Research and
  Development in Information Retrieval, ACM, pp 5--14

\bibitem[{Cormack and Grossman(2017{\natexlab{b}})}]{CormackG17Clef}
Cormack GV, Grossman MR (2017{\natexlab{b}}) Technology-assisted review in
  empirical medicine: Waterloo participation in {CLEF} ehealth 2017. In:
  Working Notes of {CLEF} 2017 - Conference and Labs of the Evaluation Forum,
  Dublin, Ireland, September 11-14, 2017.

\bibitem[{Cormack and Lynam(2005{\natexlab{a}})}]{cormack2005spam}
Cormack GV, Lynam TR (2005{\natexlab{a}}) Spam corpus creation for trec. In:
  CEAS

\bibitem[{Cormack and Lynam(2005{\natexlab{b}})}]{cormack2005trec}
Cormack GV, Lynam TR (2005{\natexlab{b}}) Trec 2005 spam track overview. In:
  TREC, pp 500--274

\bibitem[{Cormack and Mojdeh(2009)}]{cormack2009machine}
Cormack GV, Mojdeh M (2009) Machine learning for information retrieval: Trec
  2009 web, relevance feedback and legal tracks. In: TREC

\bibitem[{Cormack et~al(1998)Cormack, Palmer, and
  Clarke}]{cormack1998efficient}
Cormack GV, Palmer CR, Clarke CL (1998) Efficient construction of large test
  collections. In: Proceedings of the 21st annual international ACM SIGIR
  conference on Research and development in information retrieval, ACM, pp
  282--289

\bibitem[{Cormack et~al(2010)Cormack, Grossman, Hedin, and
  Oard}]{cormack2010overview}
Cormack GV, Grossman MR, Hedin B, Oard DW (2010) Overview of the trec 2010
  legal track. In: Proc. 19th Text REtrieval Conference, p~1

\bibitem[{Drucker et~al(2001)Drucker, Shahrary, and
  Gibbon}]{drucker2001relevance}
Drucker H, Shahrary B, Gibbon DC (2001) Relevance feedback using support vector
  machines. In: ICML, pp 122--129

\bibitem[{Dumais(2005)}]{interactive-trec-track-putting-user-search}
Dumais S (2005) The Interactive TREC Track: Putting the User Into Search. MIT
  Press,
  \urlprefix\url{https://www.microsoft.com/en-us/research/publication/interactive-trec-track-putting-user-search/}

\bibitem[{Grossman et~al(2016)Grossman, Cormack, and
  Roegiest}]{roegiest2016trec}
Grossman M, Cormack G, Roegiest A (2016) Trec 2016 total recall track overview.
  Proc TREC-2016

\bibitem[{Grossman and Cormack(2011)}]{grossman2010technology}
Grossman MR, Cormack GV (2011) Technology-assisted review in e-discovery can be
  more effective and more efficient than exhaustive manual review. Rich JL \&
  Tech 17:1

\bibitem[{Grossman et~al(2011)Grossman, Cormack, Hedin, and
  Oard}]{grossman2011overview}
Grossman MR, Cormack GV, Hedin B, Oard DW (2011) Overview of the trec 2011
  legal track. In: TREC, vol~11

\bibitem[{Hearst(2009)}]{hearst2009search}
Hearst M (2009) Search user interfaces. Cambridge University Press

\bibitem[{Hedin et~al(2009)Hedin, Tomlinson, Baron, and
  Oard}]{hedin2009overview}
Hedin B, Tomlinson S, Baron JR, Oard DW (2009) Overview of the trec 2009 legal
  track. Tech. rep., NATIONAL ARCHIVES AND RECORDS ADMINISTRATION COLLEGE PARK
  MD

\bibitem[{Hersh and Bhupatiraju(2003)}]{hersh2003trec}
Hersh WR, Bhupatiraju RT (2003) Trec genomics track overview. In: TREC, vol
  2003, pp 14--23

\bibitem[{Hogan et~al(2008)Hogan, Reinhart, Brassil, Gerber, Rugani, and
  Jade}]{hogan2008h5}
Hogan C, Reinhart J, Brassil D, Gerber M, Rugani SM, Jade T (2008) H5 at trec
  2008 legal interactive: user modeling, assessment \& measurement. Tech. rep.,
  H5 SAN FRANCISCO CA

\bibitem[{Kanoulas et~al(2017)Kanoulas, Li, Azzopardi, and
  Spijker}]{kanoulas2017clef}
Kanoulas E, Li D, Azzopardi L, Spijker R (2017) Clef 2017 technologically
  assisted reviews in empirical medicine overview. Working Notes of CLEF pp
  11--14

\bibitem[{Kolcz and Cormack(2009)}]{kolcz2009genre}
Kolcz A, Cormack GV (2009) Genre-based decomposition of email class noise. In:
  Proceedings of the 15th ACM SIGKDD international conference on Knowledge
  discovery and data mining, ACM, pp 427--436

\bibitem[{Liu and Croft(2002)}]{liu2002passage}
Liu X, Croft WB (2002) Passage retrieval based on language models. In:
  Proceedings of the eleventh international conference on Information and
  knowledge management, ACM, pp 375--382

\bibitem[{Maddalena et~al(2016)Maddalena, Basaldella, De~Nart, Degl'Innocenti,
  Mizzaro, and Demartini}]{maddalena2016crowdsourcing}
Maddalena E, Basaldella M, De~Nart D, Degl'Innocenti D, Mizzaro S, Demartini G
  (2016) Crowdsourcing relevance assessments: The unexpected benefits of
  limiting the time to judge. In: Fourth AAAI Conference on Human Computation
  and Crowdsourcing

\bibitem[{Oard et~al(2008)Oard, Hedin, Tomlinson, and Baron}]{oard2008overview}
Oard DW, Hedin B, Tomlinson S, Baron JR (2008) Overview of the trec 2008 legal
  track. Tech. rep., MARYLAND UNIV COLLEGE PARK COLL OF INFORMATION STUDIES

\bibitem[{Over(2001)}]{over2001trec}
Over P (2001) The trec interactive track: an annotated bibliography.
  Information Processing \& Management 37(3):369--381

\bibitem[{Pickens et~al(2015)Pickens, Gricks, Hardi, and
  Noel}]{DBLP:conf/trec/PickensGHN15}
Pickens J, Gricks T, Hardi B, Noel M (2015) A constrained approach to manual
  total recall. In: Proceedings of The Twenty-Fourth Text REtrieval Conference,
  {TREC} 2015, Gaithersburg, Maryland, USA, November 17-20, 2015,
  \urlprefix\url{http://trec.nist.gov/pubs/trec24/papers/catres-TR.pdf}

\bibitem[{{Rahbariasl, Shahin}(2018)}]{RahbariaslShahin2018}
{Rahbariasl, Shahin} (2018) The effects of time constraints and document
  excerpts on relevance assessing behavior. Master's thesis,
  \urlprefix\url{http://hdl.handle.net/10012/13678}

\bibitem[{Robertson and Soboroff(2002)}]{robertson2002trec}
Robertson SE, Soboroff I (2002) The trec 2002 filtering track report. In: TREC,
  vol 2002, p~5

\bibitem[{Roegiest et~al(2015)Roegiest, Cormack, Grossman, and
  Clarke}]{roegiest2015trec}
Roegiest A, Cormack G, Grossman M, Clarke C (2015) Trec 2015 total recall track
  overview. Proc TREC-2015

\bibitem[{Ruthven and Lalmas(2003)}]{ruthven2003survey}
Ruthven I, Lalmas M (2003) A survey on the use of relevance feedback for
  information access systems. The Knowledge Engineering Review 18(2):95--145

\bibitem[{Salton et~al(1993)Salton, Allan, and Buckley}]{salton1993approaches}
Salton G, Allan J, Buckley C (1993) Approaches to passage retrieval in full
  text information systems. In: Proceedings of the 16th annual international
  ACM SIGIR conference on Research and development in information retrieval,
  ACM, pp 49--58

\bibitem[{Sanderson(1998)}]{sanderson1998accurate}
Sanderson M (1998) Accurate user directed summarization from existing tools.
  In: Proceedings of the seventh international conference on Information and
  knowledge management, ACM, pp 45--51

\bibitem[{Sanderson and Joho(2004)}]{sanderson2004forming}
Sanderson M, Joho H (2004) Forming test collections with no system pooling. In:
  Proceedings of the 27th annual international ACM SIGIR conference on Research
  and development in information retrieval, ACM, pp 33--40

\bibitem[{Smucker and Jethani(2010)}]{smucker2010human}
Smucker MD, Jethani CP (2010) Human performance and retrieval precision
  revisited. In: Proceedings of the 33rd international ACM SIGIR conference on
  Research and development in information retrieval, ACM, pp 595--602

\bibitem[{Soboroff and Robertson(2003)}]{soboroff2003building}
Soboroff I, Robertson S (2003) Building a filtering test collection for trec
  2002. In: Proceedings of the 26th annual international ACM SIGIR conference
  on Research and development in informaion retrieval, ACM, pp 243--250

\bibitem[{Sparck~Jones and Van~Rijsbergen(1975)}]{spark1975report}
Sparck~Jones K, Van~Rijsbergen C (1975) Report on the need for and provision of
  an'ideal'information retrieval test collection. Computer Laboratory

\bibitem[{Tombros and Sanderson(1998)}]{tombros1998advantages}
Tombros A, Sanderson M (1998) Advantages of query biased summaries in
  information retrieval. In: Proceedings of the 21st annual international ACM
  SIGIR conference on Research and development in information retrieval, ACM,
  pp 2--10

\bibitem[{Tomlinson et~al(2007)Tomlinson, Oard, Baron, and
  Thompson}]{tomlinson2007overview}
Tomlinson S, Oard DW, Baron JR, Thompson P (2007) Overview of the trec 2007
  legal track. In: TREC

\bibitem[{Voorhees(2000)}]{voorhees2000variations}
Voorhees EM (2000) Variations in relevance judgments and the measurement of
  retrieval effectiveness. Information processing \& management 36(5):697--716

\bibitem[{Voorhees and Harman(2000)}]{Voorhees00overviewof}
Voorhees EM, Harman D (2000) Overview of the eighth text retrieval conference
  (trec-8). pp 1--24

\bibitem[{Voorhees et~al(2005)Voorhees, Harman et~al}]{voorhees2005trec}
Voorhees EM, Harman DK, et~al (2005) TREC: Experiment and evaluation in
  information retrieval, vol~1. MIT press Cambridge

\bibitem[{Wallace et~al(2010)Wallace, Small, Brodley, and
  Trikalinos}]{wallace2010active}
Wallace BC, Small K, Brodley CE, Trikalinos TA (2010) Active learning for
  biomedical citation screening. In: Proceedings of the 16th ACM SIGKDD
  international conference on Knowledge discovery and data mining, ACM, pp
  173--182

\bibitem[{Wallace et~al(2013)Wallace, Dahabreh, Schmid, Lau, and
  Trikalinos}]{wallace2013modernizing}
Wallace BC, Dahabreh IJ, Schmid CH, Lau J, Trikalinos TA (2013) Modernizing the
  systematic review process to inform comparative effectiveness: tools and
  methods. Journal of comparative effectiveness research 2(3):273--282

\bibitem[{Wang(2011)}]{wang2011accuracy}
Wang J (2011) Accuracy, agreement, speed, and perceived difficulty of users’
  relevance judgments for e-discovery. In: Proceedings of SIGIR Information
  Retrieval for E-Discovery Workshop, vol~1

\bibitem[{Wang and Soergel(2010)}]{wang2010user}
Wang J, Soergel D (2010) A user study of relevance judgments for e-discovery.
  In: Proceedings of the 73rd ASIS\&T Annual Meeting on Navigating Streams in
  an Information Ecosystem-Volume 47, American Society for Information Science,
  p~74

\bibitem[{Yu et~al(2016)Yu, Kraft, and Menzies}]{yu2016read}
Yu Z, Kraft NA, Menzies T (2016) How to read less: Better machine assisted
  reading methods for systematic literature reviews. arXiv preprint
  arXiv:161203224

\bibitem[{Zhang et~al(2015)Zhang, Lin, Wang, Clarke, and
  Smucker}]{DBLP:conf/trec/ZhangLWCS15}
Zhang H, Lin W, Wang Y, Clarke CLA, Smucker MD (2015) Waterlooclarke: {TREC}
  2015 total recall track. In: Proceedings of The Twenty-Fourth Text REtrieval
  Conference, {TREC} 2015, Gaithersburg, Maryland, USA, November 17-20, 2015

\bibitem[{Zhang et~al(2016)Zhang, Lin, Cormack, and
  Smucker}]{zhang2016sampling}
Zhang H, Lin J, Cormack GV, Smucker MD (2016) Sampling strategies and active
  learning for volume estimation. In: Proceedings of the 39th International ACM
  SIGIR conference on Research and Development in Information Retrieval, ACM,
  pp 981--984

\bibitem[{Zhang et~al(2017)Zhang, Abualsaud, Ghelani, Ghosh, Smucker, Cormack,
  and Grossman}]{zhang2017trec}
Zhang H, Abualsaud M, Ghelani N, Ghosh A, Smucker MD, Cormack GV, Grossman MR
  (2017) Uwaterloomds at the trec 2017 common core track. TREC

\bibitem[{Zhang et~al(2018)Zhang, Abualsaud, Ghelani, Smucker, Cormack, and
  Grossman}]{zhang2018cikm}
Zhang H, Abualsaud M, Ghelani N, Smucker MD, Cormack GV, Grossman MR (2018)
  Effective user interaction for high-recall retrieval: Less is more. In:
  Proceedings of the 27th ACM International Conference on Information and
  Knowledge Management, CIKM '18, pp 187--196, \doi{10.1145/3269206.3271796},
  \urlprefix\url{http://doi.acm.org/10.1145/3269206.3271796}

\end{thebibliography}


%
%

\end{document}